\documentclass[a4paper,12pt]{article}
\usepackage{epsfig}
\usepackage{amssymb}
\usepackage{graphicx}
\usepackage{color}
\definecolor{rosso}{cmyk}{0,1,1,0.4}
\definecolor{rossos}{cmyk}{0,1,1,0.55}
\definecolor{rossoc}{cmyk}{0,1,1,0.2}
\definecolor{blu}{cmyk}{1,1,0,0.3}
\definecolor{blus}{cmyk}{1,1,0,0.6}
\definecolor{bluc}{cmyk}{1,1,0,0.1}
\definecolor{verde}{cmyk}{0.92,0,0.59,0.25}
\definecolor{verdec}{cmyk}{0.92,0,0.59,0.15}
\definecolor{verdes}{cmyk}{0.92,0,0.59,0.4}

\def\circa#1{\,\raise.3ex\hbox{$#1$\kern-.75em\lower1ex\hbox{$\sim$}}\,}

\newcommand{\eV}{\,{\rm eV}}

\newcommand{\GeV}{\,{\rm GeV}}

\makeatletter
\def\art{\@ifnextchar[{\eart}{\oart}}
\def\eart[#1]#2#3#4#5#6{{\rm #2}, {\em #3 \rm #4} {\rm (#6) #5 ({\em #1})}}
\def\hepart[#1]#2{{\rm #2, \em#1}}
\newcommand{\oart}[5]{{\rm #1}, {\em #2 \rm #3} {\rm (#5) #4}}

\newcommand{\beq}{\begin{equation}}
\newcommand{\eeq}{\end{equation}}
\newcommand{\bea}{\begin{eqnarray}}
\newcommand{\eea}{\end{eqnarray}}
\newcommand{\ba}{\begin{array}}
\newcommand{\ea}{\end{array}}
\newcommand{\bi}{\begin{itemize}}
\newcommand{\ei}{\end{itemize}}
\newcommand{\bn}{\begin{enumerate}}
\newcommand{\en}{\end{enumerate}}
\newcommand{\bc}{\begin{center}}
\newcommand{\ec}{\end{center}}

\newcommand{\gsim}{\lower.7ex\hbox{$\;\stackrel{\textstyle>}{\sim}\;$}}
\newcommand{\lsim}{\lower.7ex\hbox{$\;\stackrel{\textstyle<}{\sim}\;$}}

\begin{document}
\tolerance=100000
\thispagestyle{empty}
\setcounter{page}{0}

\begin{flushright}
\texttt{CERN-PH-TH/2006-093}\\
\texttt{DNI-UAN/06/97FT}\\
\texttt{IFT-UAM/CSIC-06-23}\\
\texttt{LPT-Orsay/06-21}\\
\texttt{LYCEN 2006-07}\\
%\texttt{hep-ph/06????} 
\end{flushright}

\vspace{1cm}

\begin{center}
{\LARGE \bf 
Flavour Matters in Leptogenesis \\[0.3cm]
}
\vskip 25pt
{\bf A. Abada $^{1}$, S.  Davidson $^{2}$, A. Ibarra$^{3}$, 
F.-X. Josse-Michaux$^{1}$,\\
 M. Losada $^{4}$ and A. Riotto$^{5}$}
 
\vskip 10pt  
$^1${\it Laboratoire de Physique Th\'eorique, 
Universit\'e de Paris-Sud 11, B\^atiment 210,\\ 91405 Orsay Cedex,
France} \\
$^2${\it Institut de Physique Nucl\'eaire de Lyon, Universit\'e C-B Lyon-1,
69622 Villeurbanne, cedex,  France }\\
$^3${\it Instituto de F\'isica Te\'orica, CSIC/UAM, C-XVI,\\
Universidad Aut\'onoma de Madrid,\\ Cantoblanco, 28049 Madrid, Spain}\\
$^4${\it Centro de Investigaciones, 
Universidad Antonio Nari\~{n}o, \\ Cll. 58A No. 37-94,  Bogot\'{a},
Colombia}\\
$^5${\it CERN Theory Division, Geneve 23, CH-1211, Switzerland}

\vspace{1cm}
{\large\bf Abstract}
\end{center}
\begin{quote}
{\noindent
We give analytic approximations to the  baryon
asymmetry produced by  thermal leptogenesis  with
hierarchical right-handed neutrinos.  
Our calculation includes flavour-dependent washout processes
and CP violation in scattering, and neglects gauge
interactions and finite temperature corrections. 
Our approximate formulae 
depend upon the three CP asymmetries in the individual lepton 
flavours
as well as on three flavour-dependent efficiency factors.  
We show that the commonly used expressions for the lepton asymmetry, 
which depend on the total CP asymmetry and one single efficiency
factor, may fail to  reproduce the correct lepton asymmetry in a 
number of cases. We illustrate the importance of
using the flavour-dependent formulae in the 
context of a two right-handed neutrino model.

}

\end{quote}
\setcounter{page}{1}
%\tableofcontents

 \newpage

{\bf Note Added:} In the published version of this paper,
we applied incorrectly the $A$ matrix \cite{barbieri},
which relates the asymmetries in $B/3 - L_\alpha$ to the asymmetries
$Y_{\alpha \alpha}$ 
carried by the lepton doublets. This over-estimated the baryon
asymmetry by a factor  of order one (or a few). In this revised
version, we modify sections 4.1.4 and 4.2 to give a more correct
relation between our analytic estimates for the flavour asymmetries,
and the final baryon asymmetry.
We thank E.J. Chun and M. Plumacher for discussions
on this issue, and  in particular A Strumia for details
of his calculations and results. 

{\bf Second Note Added:} There are various other errors (we thank
S Antusch, P Di Bari,  EJ Chun, and C Gonzalez-Garcia for
comments) which we correct in this version:
\begin{enumerate}
\item In the previous net version (modified to
use more correctly the A matrix), the overall sign
of the baryon asymmetry was wrong in eqns (\ref{ppYB})
and (\ref{BAUM1large}).
 \item The coefficient  of the washout term in
the Boltzmann Equations was  a factor of 2 too large: 
in  the last term of eqns (\ref{old1}), (\ref{new1b}), 
and (\ref{newdiagonal}), and in the corresponding coefficients
of the analytic solutions, 
the numerical factor should be 1/4  (not the  1/2
of the published version).
\item The vev of the Higgs should be taken  175 GeV, not
246 GeV(given after eqn (\ref{L})). This changes $m_*$ from 
the $2-3 \times 10^{-3}$ eV used  in the published version
(at eqns (\ref{oldK}),(\ref{alphabeta})) to  $m_*  \simeq 10^{-3}$ eV. 
{\it NB:}  in the analytic approximations this 
erroneous factor of 2 cancels against the
one  of point 2 above. So the analytic
approximations  remain acceptable , using  $m_*  \simeq 10^{-3}$ eV.
\item Factors of two in eqns (\ref{later}), (\ref{bd})
and (\ref{hateps})  have been corrected.
%\item  In the published version, 
%the captions of  Figures 1,2, and 3 
%indicate  that  $e$  was taken distinguishable  from $\mu$.
%However, for    $M_1 > 10^9$ GeV,
%only the $\tau$ is distinguishable \footnote{unless
%tan $\beta \gsim 5$ in a  SUSY model}.  We
%therefore interchanged   the $e$ and $\tau$ indices
%on the inputs ($\epsilon_{ee} \rightarrow \epsilon_{\tau \tau}$,
%$K_{\tau\tau} \rightarrow K_{ee}$, ...). We also
%changed a washout factor in Figure 2 to account
%for points 2 and 3 above.  We have remade the
%plots
\item  In the published version, 
the captions of  Figures 1,2, and 3 
indicate  that    $e$  was taken
distinguishable from $\mu$.
However, for    $M_1 > 10^9$ GeV,
only the $\tau$ is distinguishable\footnote{unless
tan $\beta \gsim 5$ in a  SUSY model}.  We
therefore interchanged   the $e$ and $\tau$ indices
on the inputs ($\epsilon_{ee} \rightarrow \epsilon_{\tau \tau}$,
$K_{\tau\tau} \rightarrow K_{ee}$,...). We also
changed a washout factor in the caption of
 Figure 2 to account
for points 2 and 3 above.   We have remade
the plots, with these modified inputs, and
the plots remain unchanged.

 \end{enumerate}

\section{Introduction}
There are robust  observational evidences that a tiny excess
of matter over antimatter was produced
in our Universe \cite{sakharov}, but its origin is still a mystery.
Baryogenesis through Leptogenesis \cite{FY} is 
a simple  mechanism to 
explain this baryon asymmetry
of the Universe. A lepton asymmetry is dynamically generated
and then  converted into a baryon asymmetry
due to $(B+L)$-violating sphaleron interactions \cite{kuzmin}
which exist in the Standard Model (SM).

A simple model in which this mechanism can be implemented is ``Seesaw''(type I)
\cite{seesaw}, consisting
of the Standard Model (SM)
plus two or three   right-handed (RH) Majorana neutrinos.
In this simple extension of the SM, the usual scenario 
that is explored  (referred to as
``thermal leptogenesis'') consists of
a hierarchical spectrum for the RH neutrinos, such that the 
lightest of the RH neutrinos
is produced by thermal scattering after inflation,  and subsequently
decays out-of-equilibrium in a lepton number and CP-violating way, 
thus satisfying Sakharov's constraints.  

In recent years, a lot of 
work \cite{lept,ogen,work}, 
has been devoted to a thorough analysis of this model,
giving limited attention to
the issue of lepton flavour  \cite{barbieri}.  The    dynamics of
leptogenesis is usually  addressed within the `one-flavour' approximation,
where  Boltzmann equations
are written for the abundance of the 
lightest RH neutrino, responsible
for the out of equilibrium and CP asymmetric decays, 
and  for the total lepton asymmetry. However, this 
`one-flavour' approximation is
rigorously correct only when the interactions mediated by 
charged lepton Yukawa couplings are out of equilibrium.

In ref. \cite{barbieri}, flavoured Boltzmann Equations were written down. %charged lepton Yukawa couplings were  included 
%in thermal leptogenesis  focusing on their effect
%in  $\Delta L = 2$ scattering. 
Flavour effects in ``resonant leptogenesis'' were
studied in \cite{PU},   discussed for thermal 
leptogenesis in  the two-right-handed
neutrino model in  \cite{EMX}, and
used in \cite{oscar} to protect an asymmetry made in
the decay of the middle right-handed neutrino.
 In the four generation models of \cite{AM}, flavour
was used to enhance the asymmetry, foreshadowing
the results we obtain here.
The impact of  flavour in thermal leptogenesis has been recently 
studied in some detail 
\cite{davidsonetal,nardietal}, including the
quantum oscillations/correlations of the asymmetries in lepton flavour space 
\cite{davidsonetal}. It was shown that 
the Boltzmann equations describing the asymmetries in flavour space 
have  additional terms which can  significantly affect the result for the
final baryon asymmetry. In
\cite{davidsonetal}, we focused 
on how flavour effects can enlarge the
area of parameter space where leptogenesis can work:
 the lower bound on the reheat
temperature is mildly decreased, and  the upper bound
on the light neutrino mass scale  no longer 
holds\footnote{The bound is removed when flavour effects
are relevant, which is the case  for leptogenesis at temperatures
$\lsim 10^{12}$ GeV.}.

Flavour effects have not usually been included in leptogenesis
calculations.  This is perhaps because perturbatively, they seem to be
a small correction. For instance, if the asymmetry is a consequence of the
very-out-of-equilibrium decay of an initial population of
right-handed neutrinos, then  the total lepton
asymmetry is of order $\epsilon/g_*$, where $\epsilon$
is the total CP asymmetry in the decay, and $g_*$ counts for 
the entropy dilution factor. Clearly the small
charged lepton Yukawa couplings have no effect on $\epsilon$.
 However,
realistic leptogenesis is a drawn-out dynamical
process, involving  the  production and 
destruction of right-handed neutrinos, 
and  of a  lepton asymmetry  that is distributed among
{\it distinguishable} flavours.  The processes which
wash out lepton number
are flavour dependent, {\it e.g} the inverse decays
from electrons can destroy the lepton asymmetry carried by,
and only by,  the electrons.
The  asymmetries  in each flavour
are  therefore washed out differently,
and will appear with different weights in the final formula
for the baryon asymmetry.  This is physically inequivalent
to the treatment of washout in the one-flavour approximation,
 where  
%the flavours are taken indistinguishable, 
indistinguishable 
 leptons propagate between decays and inverse decays,
so inverse decays  from all flavours are taken to wash out asymmetries
in any flavour\footnote{The ``one-flavour'' formulae
describe  leptogenesis  that takes place at
temperatures larger than $10^{12}$ GeV, before the charged lepton Yukawas
come into equilibrium.  They are also appropriate for
right-handed neutrinos who decay only to one flavour.
(But note from  eqns (\ref{third},\ref{interpolate}), that flavour effects can
be important when there are small branching ratios 
to other flavours.)}.

In this paper we  provide the necessary
analytical expressions for the computation of the baryon asymmetry
including flavour that the interested reader may apply to their preferred 
model. By comparing  to  the usually adopted `one-flavour' 
approximation, 
we will show that the commonly used expressions for the lepton asymmetry, 
which depend on the total CP asymmetry and one single efficiency
factor, fail to  reproduce the correct lepton asymmetry in a large
number of cases. 
As an application, we also present in this paper a detailed analysis of flavour effects
on lepton asymmetries for a two right-handed neutrino model. 
Explicit examples in which sizeable enhancements
can be obtained are also given.

The paper is organized as follows. In Section 2, we review the 
conventional flavoured-blind computation of the baryon asymmetry, 
and present some analytic approximations that will be used later. 
In Section 3,  we introduce the Boltzmann equations  we will
solve, which differ by the inclusion of flavour and
of CP violation in $\Delta L = 1$ processes.
The following section
provides a list of rules and expressions 
to apply in order to obtain an estimate  of
the baryon asymmetry which includes flavour effects. 
Section 5 contains the analysis in the 
context of two right-handed neutrino model which make manifest the difference 
between the results when flavours are and are not included. 
In section 6 different textures for
the neutrino Yukawa coupling matrix and their implications are explored.
In Section 7 we discuss the special case in which there is no
CP violation in the right-handed neutrino sector, and finally
in Section 8 we draw our conclusions. 

\section{The conventional computation of the baryon asymmetry}

This section introduces notation and reviews  the 
calculation of the lepton asymmetry when 
the charged lepton Yukawa couplings are neglected. 
As we shall see, the commonly used formulae for the final lepton
asymmetry, which  we report in this section,
may not  be appropriate   once flavours
are considered.

Our starting point is the Lagrangian of the Standard Model (SM)
with the addition of three right-handed neutrinos $N_{i}$ ($i=1,2,3$)
with heavy Majorana masses $M_{3}> M_{2}> M_{1}$
and Yukawa couplings $\lambda_{i\alpha}$. Working in
the basis in which the Yukawa couplings for the
charged leptons are  diagonal, the Lagrangian reads
\begin{equation}
\label{L}
{\cal L} = {\cal L}_{\rm SM} + \left(\frac{M_{i}}{2} N_i^2 + 
\lambda_{i\alpha} N_i  \ell_\alpha \, H + h_\alpha\,
H^c \,\bar{e}_{R \alpha} \ell_\alpha +\hbox{h.c.}\right)\, .
\end{equation}
Here $\ell_\alpha$ and $e_{R\alpha}$ indicate the lepton doublet and
singlet with flavour 
$(\alpha=e,\mu,\tau)$ respectively,  and $H$ is the Higgs doublet whose
neutral component has a vacuum expectation value equal to
$v=175$ GeV.

We assume that right-handed neutrinos are hierarchical,
$M_{2,3}\gg M_1$ so that studying the 
evolution of
the number density of  $N_1$ suffices.  The 
final amount of $({\cal B}-{\cal L})$ asymmetry can be parametrized
as 
$Y_{{\cal B}-{\cal L}}=n_{{\cal B}-{\cal L}}/s$, 
where $s=2\pi^2 g_* T^3/45$ is the entropy density and
$g_*$ counts the effective
number of
spin-degrees of freedom in thermal equilibrium ($g_*=217/2$
in the SM
with a single generation of right-handed
neutrinos).  After reprocessing by
sphaleron 
transitions, the baryon asymmetry is related to the ${\cal L}$
asymmetry by \cite{Harvey:1990qw}
\begin{equation}\label{eq:YB}
Y_{\cal B}=-\left( \frac{8 n_G+4n_H}{14n_G+9n_H}\right)Y_{\cal L},
%Y{\cal B}{s}=\frac{24+4n_H}{66+13n_H}\, Y_{{\cal B}- {\cal L}}\, ,
\end{equation}
where $n_H$ is the number of Higgs doublets, and $n_G$ 
the number of fermion generations (in equilibrium).

One  defines the CP asymmetry 
generated by $N_1$ decays  as

\begin{equation}
\epsilon_{1} \equiv 
\frac{\sum_\alpha[\Gamma(N_1\rightarrow H \ell_\alpha)-\Gamma(
N_1\rightarrow \overline{H} \overline{\ell}_\alpha)]}{
\sum_\alpha[\Gamma(N_1\rightarrow H
\ell_\alpha)+\Gamma
(N_1\rightarrow \overline{H} \overline{\ell}_\alpha)]}=
\frac{1}{8\pi}\sum_{j\neq 1}
\frac{\textrm{Im}
\left[ (\lambda \lambda^{\dagger})_{j 1}^2\right] }{\left[\lambda
\lambda^{\dagger}\right]_{11}}
g\left(\frac{M_{j}^2}{M_{1}^2}\right)\, ,
\end{equation}
where  the wavefunction plus vertex contributions are included in
\cite{Covi:1996wh}
\begin{equation}
g(x)=\sqrt{x}\left[ \frac{1}{1-x} + 1 -(1+x)\ln
\left(\frac{1+x}{x} \right)
 \right] \stackrel{x\gg 1}{\longrightarrow} - \frac{3}{2
\sqrt{x}} \, .
\end{equation}
Notice, in particular, that $\epsilon_1$ denotes
the CP asymmetry in the total number (the trace) of flavours.

Besides  the CP parameter $\epsilon_1$, the final baryon asymmetry
depends on a single wash-out parameter,

\begin{equation}
\label{oldK}
K\equiv\frac{\sum_\alpha\Gamma(N_1\rightarrow
H\ell_\alpha)}{H(M_1)}\equiv
\left(\frac{\widetilde{m}_1}{\widetilde{m}^*}\right )\, ,
\end{equation}
where $H(M_1)$ denotes the value of the Hubble rate
evaluated at a temperature $T=M_1$ ($\widetilde{m}^*\sim  10^{-3}\,
{\rm eV}$) and 

\begin{equation}
\tilde{m}_1\equiv \frac{(\lambda \lambda^\dagger)_{11}
v^2}{M_{1}}\, 
\end{equation} 
is proportional to
the total decay rate of the right-handed neutrino $N_1$. 
 
By defining the variable $z=M_1/T$,
the Boltzmann equations for  the lepton asymmetry $Y_{\cal L}$,
and the  right-handed neutrino number density $Y_{N_1}$
(both normalised to the entropy $s$),  may be written
in a compact form as

\beq
 \frac{d(Y_{N_{1}} -Y_{N_{1}}^{\rm EQ} )}{dz} = -  \frac{z}{s H(M_1)} \left( \gamma_D + \gamma_{\Delta L =1}\right)
 \left(\frac{Y_{N_{1}}}{Y_{N_{1}}^{\rm EQ}} -1\right) -
\frac{d Y_{N_{1}}^{\rm EQ} }{dz}\,  ,
\label{newold1}
\eeq

\beq
\frac{dY_{\cal L}}{dz}  =    \frac{z}{s H(M_1)}\left[\left(
\frac{Y_{N_{1}}}{Y_{N_{1}}^{\rm EQ}} - 1\right)\epsilon_1 \gamma_D -
\frac{Y_{\cal L}}{Y_L^{\rm EQ}} \left(\gamma_D
+\gamma_{\Delta L =1} + \gamma_{\Delta L=2}\right)\right]\, .
\label{newold2}
\eeq
The processes taken into account in
these equations  are decays and  inverse
decays with rate $\gamma_D$,  $\Delta L=1$ scatterings
such as $(q t^c \rightarrow N \ell)$, 
and $\Delta L=2$ processes mediated
by heavy neutrinos. The first three modify the abundance of the
lightest right-handed neutrinos.  The  $\Delta L=2$
scatterings mediated by $N_{2,3}$ are  
neglected  in our analysis for 
 simplicity\footnote{See,{\it e.g.} the Appendix of \cite{davidsonetal}.
We discuss later the restrictions this implies.}. 
The  various $\gamma$
are thermally averaged rates,
 including all contributions
summed over flavour ($s$, $t$ channel
interference etc);  explicit expressions can be
found in the literature (see for example   \cite{lept,ogen}).
Notice that in this ``usual'' analysis,  $\Delta L=1$ scattering contributes
to the creation of $N_1$'s and not to the production of a lepton
asymmetry, only to the washout
\footnote{We thank A Strumia, A. Pilaftsis, G. Giudice and 
E. Nardi for useful conversations about this point.}. 
This is a minor point in the single flavour analysis;
it is more relevant when flavour is included, and will
be discussed  in the following two sections.

Approximate analytic solutions for 
$Y_{\cal L}$ and $\Delta_{N_1}\equiv Y_{N_1}-Y^{\rm EQ}_{N_1}$,
 which reproduce
the numerical plots \cite{lept,ogen}, can be obtained 
from  simplified equations \cite{barbieri,ogen}.
Calculating in zero temperature field theory for 
simplicity\footnote{Significant finite temperature
corrections were found in \cite{lept}, which have ${\cal O}(1)$ effects on
the final asymmetry. }, one obtains
\beq
\gamma_D \simeq s Y_{N_1}^{EQ} \frac{K_1(z)}{K_2(z)} \Gamma_D, ~~~~
Y^{\rm EQ}_{N_1} \simeq \frac{1}{4 g_*}z^2 \,K_2(z)\,.
\eeq
The Boltzmann equations can be approximated
\begin{eqnarray}
\label{old2}
\Delta^\prime_{N_1}&=&- z K \frac{K_1(z)}
{K_2(z)}\, f_1(z) \,\Delta_{N_1}-Y^{\rm EQ\,\prime}_{N_1}\, ,\,\,
 \\
\label{old1}
Y^\prime_{\cal L}&=&\epsilon_1    K z\, \frac{K_1(z)}
{K_2(z)}
\Delta_{N_1}-\frac{1}{4}z^3\,K\,K_1(z)\, f_2(z)\,\, Y_{\cal L}\, 
\end{eqnarray}
where
$K_1$ and $K_2$ are  modified Bessel functions of the second kind.
The  function $f_1(z)$ accounts for the 
presence of $\Delta L=1$ scatterings \cite{lept,ogen},
and $f_2(z)$ accounts for scatterings in the washout
term of the asymmetry. They can be approximated
\cite{ogen},
in interesting limits, as 
\begin{equation}
f_1(z) \simeq  
\left\{ \begin{array}{ccc} 1 & &  {\rm for}~ z\gg 1 \\
 \frac{N_c^2 m_t^2}{ 4 \pi^2 v^2 z^2}\  & & {\rm for} ~ z\lsim 1\, ,
\end{array} \right.
\end{equation} 
and
\begin{equation}
f_2(z) \simeq  
\left\{ \begin{array}{ccc} 1 & &  {\rm for}~ z\gg 1 \\
 \frac{a_K N_c^2 m_t^2}{ 8 \pi^2 v^2 z^2} & & {\rm for} ~ z\lsim 1\, ,
\end{array} \right.
\end{equation} 
where 
$\frac{N_c^2 m_t^2}{ 8 \pi^2 v^2}\equiv K_s/K\,  \sim 0.1
$ parametrizes the strength of
 the $\Delta L=1$ scatterings and $a_K=4/3\,(2)$ for the weak (strong)
wash out case.  A
 good approximation to the rate $K z (K_1(z)/K_2(z)) f_1(z)$ is given
by the function $(K_s+K z)$ \cite{lept,ogen} while the wash out
term $-(1/4)z^3 K K_1(z) f_2(z)Y_{\cal L}$ is well approximated at
small $z$ by $-a_K K_s Y_{\cal L}$.

In the strong wash-out regime, the parameter $K\gg 1$ and the right
handed neutrinos $N_1$'s are nearly in thermal equilibrium. Under these
circumstances, one can set $\Delta^\prime_{N_1}\simeq 0$ and
$\Delta_{N_1}\simeq (z K_2/4 g_* K)$. Exploiting a saddle-point approximation
in Eq. (\ref{old1}) one easily reproduces the
fit to the numerical results \cite{lept,ogen}
\begin{equation}
\label{strongold}
Y_{\cal L}\simeq 0.3\, \frac{\epsilon_1}{g_*}\left(\frac{0.55\times
10^{-3}\,{\rm eV}}{\tilde{m}_1}
\right)^{1.16}\, .
\end{equation}

In the opposite weak wash-out regime,  assuming that no 
right-handed neutrinos are initially present in the plasma, 
there could be a cancellation in the final
lepton asymmetry between the
(anti-) asymmetry
generated  in $N_1$ production, and the
lepton asymmetry produced as the $N_1$ decay.
However, this cancellation does not occur, 
in  Eqs. (\ref{newold1}) and (\ref{newold2}),
 because  CP violation in the 
$\Delta L=1$ scatterings is not included.
$\Delta L=1$ processes   contribute significantly to the production
of  right-handed neutrinos, without making any
associated (anti-) lepton asymmetry, and the 
$N_1$  later produce a
lepton asymmetry in decay. The number of $N_1$ produced
is $\propto K$,   and the
final lepton asymmetry can be approximated
\cite{lept,ogen}
\begin{equation}
\label{weakold}
Y_{\cal L}\simeq 
0.3 \,\frac{\epsilon_1}{g_*}\left(\frac{\tilde{m}_1}{3.3\times
10^{-3}\,{\rm eV}}\right)\, .
\end{equation}
Notice  that the final baryon asymmetry in the 'one flavour
approximation' depends always upon the trace of the CP asymmetries
over flavours, $\epsilon_1$, times  a function of the trace over flavours
of the decay rate of the right-handed neutrinos, $K$. This is due
to the fact that the inverse decay  term in Eq. (\ref{newold1})
is proportional to the trace over the flavours 
of the lepton asymmetry times the trace over the flavours of the
decay rate of $N_1$'s. 
The reader is invited
to remember this point in the following when we  explain why
the `one flavour approximation' fails to predict the exact baryon
asymmetry.

\section{Including flavours and CP violation in scattering }

In this section,  we introduce the  Boltzmann
equations for individual flavour asymmetries, \cite{barbieri}. 
We define
 $Y_{\alpha\alpha}$ to be the lepton asymmetry
in flavour $\alpha$, 
where  the $\alpha$  are the  lepton mass eigenstates
at the temperature of leptogenesis. 
As discussed in \cite{davidsonetal},
the  $Y_{\alpha \alpha}$ are the diagonal elements
of a matrix $[Y]$ in flavour space,    whose trace  is the total 
lepton asymmetry.  In this paper,
the  off-diagonal elements  are neglected \footnote{See
\cite{davidsonetal} for a discussion.
% the
%off-diagonals   can encode quantum correlations.
The equations of motion  for the matrix $[Y]$  are more
complicated than the Boltzmann equations, but
at most temperatures are equivalent to 
 Boltzmann equations  written in the
mass eigenstate basis of the leptons in the plasma. 
The off-diagonal elements of $[Y]$ could have some effect
on the lepton asymmetry,  if leptogenesis takes place
just as a charged lepton Yukawa coupling is coming into
equilibrium (so the mass eigenstate basis is
changing).}.

The mass eigenstates for the particles in the Boltzmann
equations(BE)  are determined by the interactions
which are fast compared to those  processes included in the BE.  The interaction
rate for Yukawa coupling $h_\alpha$ can be estimated
as \cite{CKO}
\begin{equation}
\Gamma_{\alpha}
\simeq 5\times 10^{-3} h_\alpha^2\,T,
\end{equation}
so interactions involving
the tau  (mu) Yukawa coupling
are out-of-equilibrium in the primeval plasma if $T \gsim 10^{12}$ GeV
($T \gsim 10^{9}$ GeV) \footnote{The electron Yukawa
coupling mediates interactions relevant in the early Universe only for
temperatures beneath $\sim 10^5$ GeV and can be safely disregarded.}.
Thermal leptogenesis takes place at temperatures on the
order of $M_1$, and the asymmetry is generated when the rates
$\lsim H$, so we conclude that the $\tau$ ($\mu$)  lepton
doublet is a distinguishable mass eigenstate, for the purposes
of leptogenesis, at $ T < 10^{12} (10^9) $ GeV.

The Boltzmann equations  that we will use in
this paper, for the flavour
asymmetries $Y_{\alpha\alpha}$, are listed below.
They differ  from those of
\cite{davidsonetal} in two respects.

First, we  have neglected the off-diagonal
terms of the matrix $[Y]$. The second and most significant
difference is that  we have included
CP violation in the $\Delta L = 1$ scattering rate,  
which  will give  $Y_L \propto K^2$  for weak washout
(instead of $K$ in  eq. (\ref{weakold})). 
That is, the function $\gamma_{\Delta L=1}$, which appears in
the $N_1$ creation term, also
now appears in the first term of equation
(\ref{new2a}), which describes the production of
the lepton flavour asymmetry. 
 Later in  this section we will
calculate the CP asymmetry in scattering, and show
that it gives the same $\epsilon$ as decay and inverse decay,
in the limit of hierarchical right-handed neutrinos.
As in \cite{davidsonetal}, we continue
to neglect the non-resonant contribution to 
$\Delta L = 2$ scatterings, and its
associated flavour effects \cite{barbieri}. 
At the end of section 4.1.2, we discuss the parameter
range where this is acceptable.

Equation (\ref{newold1}) for the $N_1$ number density
 remains unchanged, and   the equation
for the flavoured lepton asymmetry is
\beq
\frac{dY^{\alpha \alpha}}{dz}  =    \frac{z}{s H(M_1)}\left[\left(
\frac{Y_{N_{1}}}{Y_{N_{1}}^{\rm EQ}} - 1\right)
\epsilon^{\alpha \alpha} (\gamma_D +\gamma_{\Delta L =1}) -
\frac{Y^{\alpha \alpha}}{Y_L^{\rm EQ}} \left(\gamma_D^{\alpha \alpha}
+\gamma_{\Delta L =1}^{\alpha \alpha} %+ \gamma_{\Delta L=2}
\right)\right]\, ,
\label{new2a}
\eeq
where there is no sum over $\alpha$ in the last term of 
equation (\ref{new2a}) (or (\ref{new1b})).

The rates  \cite{ogen} and asymmetries
are calculated in zero temperature field theory,
and include processes mediated by the neutrino
and top Yukawa couplings. This is simple, and parametrically consistent.
However,  finite temperature and gauge corrections can
be significant \cite{lept}, and should in principle
be included.

 To obtain  analytic solutions we simplify
this, with the approximations introduced in the
previous section, to

\begin{eqnarray}
\label{new1b}
Y^\prime_{\alpha \alpha}&=&\epsilon_{\alpha \alpha}K z\, \frac{K_1(z)}
{K_2(z)}f_{1}(z)\Delta_{N_1}-\frac{1}{4}z^3 K_1(z)\, f_{2}(z)\,
K_{\alpha \alpha}Y_{\alpha \alpha} \\
\label{new2b} 
\Delta^\prime_{N_1}&=&- z\, K\, \frac{K_1(z)}
{K_2(z)}\, f_1(z)\, \Delta_{N_1}-Y^{\rm EQ\,\prime}_{N_1}\, ,
\end{eqnarray}
%where there is no sum on $\alpha$ in the last term of
%(\ref{new2}), and
where 
\begin{equation}
K_{\alpha\alpha} = K \frac{\lambda_{1\alpha} 
\lambda_{1\alpha}^*}{\sum_\gamma|\lambda_{1\gamma}|^2}=\left(
\frac{\tilde{m}_{\alpha\alpha}}{ 10^{-3}\,{\rm eV}}\right)
\,  ,\,\,K=\sum_\alpha
K_{\alpha\alpha}\, . 
\label{alphabeta}
\end{equation}
$K_{\alpha\alpha}$ parametrizes the  decay rate of $N_1$ to the $\alpha$-th 
flavour, and the trace  $\sum_\alpha K_{\alpha\alpha}$, coincides with the
$K$ parameter defined in the previous section, see Eq. (\ref{oldK}).

Notice in particular that the dynamics of the right-handed neutrinos
is always set by the total $K$.

The CP asymmetry in the $\alpha$-th flavour  is  $\epsilon_{\alpha\alpha}$
and  is normalised by the total decay rate 
\begin{eqnarray}
\epsilon_{\alpha\alpha}  &  =& 
\frac{1}{(8\pi)}\frac{1}{  [\lambda \lambda^{\dagger}]_{11}}
\sum_j {\rm Im}\, \left\{ ( \lambda_{1 \alpha} ) (\lambda
\lambda^{\dagger})_{1j}
 \lambda^*_{j \alpha} \right\}
g\left(\frac{M_{j}^2}{M_{1}^{2}}\right)\, 
\label{flavour-CPasym} \\
& \rightarrow & \frac{3}{(16\pi) [\lambda \lambda^{\dagger}]_{11}}
 {\rm Im}\, \left\{ \lambda_{1 \beta} 
\frac{[m^*]_{\beta \alpha}}{v^2}\lambda_{1 \alpha}  \right\}
\label{later}
\end{eqnarray}
where the second line is in the limit of hierarchical $N_J$, and
$m   =  U^* D_m U^\dagger = v^2 \lambda^T M^{-1} \lambda$ is the light
neutrino mass matrix.  If 
$m_{max}$ is the heaviest light neutrino mass
($= m_{atm}$ for the non-degenerate case) 
 and  we  define $\epsilon^{\rm max}$
$= 3 \Delta m^2_{atm} M_1/(8 \pi v^2 m_{max})$ \cite{bound},
 then
the flavour dependent CP asymmetries are bounded  by 

\beq
\epsilon_{\alpha \alpha}  \leq \frac{ 3 M_1 m_{max}}{16 \pi v^2}
\sqrt{\frac{K_{\alpha \alpha}}{K}} = \epsilon^{\rm max} \frac{m_{max}^2}{\Delta m^2_{atm}}
 \sqrt{\frac{K_{\alpha \alpha}}{K}}
\label{bd}
\eeq
so the maximum   CP  asymmetry in a given flavour 
is unsuppressed for degenerate light neutrinos \cite{davidsonetal},
but decreases as the square root  of  the branching
ratio to that flavour $ = K_{\alpha \alpha}/K$.

The CP asymmetry $\epsilon_{\alpha\alpha}$ can be written in terms
of the diagonal matrix of the light neutrino mass eigenvalues $m={\rm 
Diag}(m_1,m_2,m_3)$, the diagonal matrix of the the right
handed neutrino masses $M={\rm Diag}(M_1,M_2,M_3)$ 
and an orthogonal complex matrix
$R=v\,M^{-1/2}\,\lambda\,U\, m^{-1/2}$ 
\cite{Casas:2001sr}, where $U$ is the leptonic
mixing matrix,  which ensures that the correct low-energy parameters are obtained

\begin{equation}
\epsilon_{\alpha\alpha}= -\frac{3 M_1}{16\pi v^2}\, \frac{{\rm Im}\left(
\sum_{\beta\rho} m_\beta^{1/2}m_\rho^{3/2} U^*_{\alpha\beta}U_{\alpha\rho}
R_{\beta 1}R_{\rho 1}\right)}{\sum_\beta m_\beta\left|R_{1\beta}\right|^2}.
\end{equation}
As noted in \cite{nardietal},  for a  real  $R$ matrix, 
the individual CP asymmetries 
$\epsilon_{\alpha\alpha}$ may not vanish because of the presence
of CP violation in the $U$ matrix. On the contrary, the total
$CP$ asymmetry 
 $\epsilon_1 = \sum_\alpha  \epsilon_{\alpha\alpha}$ vanishes. 

\subsection{The CP asymmetry in $\Delta L = 1$ scattering}

We now wish to show that the CP asymmetry in scattering
processes such as $(q t^c \rightarrow N \ell_\alpha)$  is
the same as in decays and inverse decays \footnote{We thank E. Nardi
for discussions  of his work in progress.}. This result
 was found in \cite{PU,PUb}, for the case of resonant leptogenesis.
CP violation in scattering is usually neglected in thermal
leptogenesis \cite{lept,ogen}, because observed neutrino masses favour the
strong washout regime $K >1$.  In strong washout, any contribution
to the lepton asymmetry by scattering processes  during
$N_1$ production is rapidly washed out, so the CP violation
in these processes can be neglected, for lack of 
observable consequences.  However, we wish to include
 this CP violation, because washout in one flavour could
be small, even though $K \gg 1$ and we wish to correctly
include the contribution of weakly washed out lepton flavours to
the final lepton asymmetry.

For simplicity, we work at zero temperature, in the limit
of hierachical right-handed neutrinos.  This means we calculate
in an effective field theory  with particle content of the SM $+ N_1$, and
the effects of the  heavier $N_2$ and $N_3$ appear in a
dimension five  operator $(HL_\alpha) (HL_\beta)$. For
computing  one-loop CP violating effects involving $N_1$,
we can take the coefficient of this operator 
$\propto [m_\nu]_{\alpha \beta}/v^2$.

 We define 
the CP asymmetries in 
$\Delta L = 1$ scattering (mediated
by $s$ and $t$-channel 
Higgs boson exchange) as
\beq
\hat{\epsilon}_s^{\alpha \alpha} = \frac{ 
\sigma (t^c q \rightarrow NL_{\alpha})
- \bar{\sigma} (\bar{q} \overline{t^c} \rightarrow N\bar{L}_{\alpha})}
{ \sigma + \bar{\sigma}} 
\eeq
\beq
\hat{\epsilon}_t^{\alpha \alpha} = \frac{ 
\sigma (q  N \rightarrow  \overline{t^c}L_{\alpha})
- \bar{\sigma} (  \bar{q} N  \rightarrow   t^c \bar{L}_{\alpha})}
{ \sigma + \bar{\sigma}}
= \frac{ 
\sigma (q  \bar{L}_{\alpha} \rightarrow  \overline{t^c}N )
- \bar{\sigma} (  \bar{q}L_{\alpha}  \rightarrow   t^cN )}
{ \sigma + \bar{\sigma}}
\label{?t}
\eeq
where barred fields are the antiparticles.
The initial state density factors
cancel in the ratio,  so  the cross-sections
$\sigma$, $\bar{\sigma}$ can be replaced by the matrix
elements squared $| {\cal M} |^2$, integrated over 
final state phase space $\int d \Pi$. 
If the tree + loop matrix element  is
separated into a coupling constant
part $c$ and an amplitude ${\cal A}$:
\beq
{\cal M} = c^t {\cal A}^t + c^\ell {\cal A}^\ell \, ,
\eeq
where the matrix element for the CP conjugate process
is $
\bar{ {\cal M}} = c^{t*} {\cal A}^t + c^{\ell *} {\cal A}^\ell$, 
then the CP asymmetry can be written
\beq
{\epsilon} =\frac{2  {\rm Im}\{ c^tc^{l*} \}}{  | c^{t}|^2}
 \frac{ \int {\rm Im}\{  {\cal A}^t {\cal A}^{l*} \}  d\Pi }
{ \int | {\cal A}^t|^2 d\Pi }\, .
\label{hateps}
\eeq
The  loop amplitude has an  imaginary part when
there are branch cuts corresponding to intermediate on-shell
particles, which  can arise here in a bubble 
 on the $N$ line at the
$NHL_\alpha$ vertex, {\it e.g.} for
$s$-channel Higgs exchange:
\beq
  {\rm Im}\{  {\cal A}^t {\cal A}^{l*} \}   = 
  {\cal A}^t (t^c q \rightarrow NL_{\alpha})  
\int  {\cal A}^{t*} ( t^c q \rightarrow L_{\alpha} L'_{\beta} H' )
 d\Pi'
 {\cal A}^{t*} ( L'_{\beta} H' \rightarrow N )\, .
%\propto p_L \cdot p_L' d \Pi'
\eeq
$H'$ and $L'_{\beta}$ are the (assumed massless) 
intermediate on-shell particles, and $d \Pi'$ is 
the integration over their phase space. 

In the scattering process
$c^t =  h_t \lambda^*_{1 \alpha}$ and   
$c^{\ell}  = 3  h_t   \lambda_{1 \beta}[ m^*]_{\beta \alpha} /v^2$,
where $h_t$ is the top Yukawa coupling,  
%and in the limit of  hierarchical
%right-handed neutrinos,  we replace the $N_{2,3}$ propagator
%by 
so the complex coupling constant combination in
$\hat{\epsilon}_{\alpha \alpha}(A\bar{H} \rightarrow NL_{\alpha}) $
is clearly the same as  in
$\epsilon_{\alpha \alpha}$ of eq. (\ref{later})\footnote{In both cases, there
is an overall factor of three, from the  weak $SU(2)$
index contractions. This can be seen  by
reinstating the $N_{2,3}$ propagators:
the charged and neutral component 
of the intermediate $H'$ and $L'$  
contribute  in the  $N_1$ wave-function correction,
giving a 2, but only  the charged or
the neutral $H'$ and $L'$  appear in the
vertex correction.}.
%The  $| {\cal A}^t(A\bar{H} \rightarrow NL_{\alpha})|^2$
%downstairs in eqn (\ref{hateps}) is $\propto  p_N \cdot p_L$
To obtain the amplitude ratio
(the  second fraction in 
eqn (\ref{hateps})), we take, for  instance
$ {\cal A}^t (N \rightarrow \bar{H}\bar{L^\alpha}) = 
\bar{u}_\ell P_L u_N$, and after straightforward  spin sums, 
one sees that  it  is 
the same for scattering and $N$ decay,  so
$\hat{\epsilon}_s^{\alpha \alpha} = \hat{\epsilon}_t^{\alpha \alpha} = \epsilon^{\alpha \alpha}$.

 The equality (\ref{?t}), and the result that 
the $s$ and $t$-channel  scattering asymmetries are
equal to the decay asymmetry, may be an artifact of
our effective field theory calculation, where  there
is no  momentum 
exchanged on the $N_{2,3}$ line in the loop.

In the finite temperature calculation of \cite{lept},
 the Higgs boson has a large thermal mass due to its interactions 
with the top quarks.  At $T \gg M_1$ it can therefore decay to
$N L$, and ref.  \cite{lept} found a CP asymmetry in this decay. 
Some part of the $N$ production that is included in zero
temperature  scattering
computations, is resummed in the finite temperature
$H$ decays,  so it is consistent that in both processes, 
a lepton asymmetry can be generated.

\subsubsection*{Checking CPT and unitarity}

We would like to verify that the  CP violating matrix elements 
${\cal M}$ for   $\Delta L = 1$
scattering processes,  satisfy  the constraints
following from CPT invariance and $S$-matrix unitarity.
See {\it e.g.} \cite{Covi} for useful related discussions.
For $i$  any state, and $\bar{i}$ its CP
conjugate,  CPT and unitarity imply \cite{Kolb}
\beq
\sum_{\bar{j}}|{\cal M}(\bar{i} \rightarrow \bar{j})|^2 = 
\sum_{\bar{j}}
 |{\cal M}( \bar{j} \rightarrow \bar{i} )|^2 = 
\sum_{j}
|{\cal M}(i \rightarrow j)|^2,
\label{CPTunit}
\eeq
where $\{ j \}  $ is the set of
accessible states.
We want to check that this is consistent with
having a CP asymmetry in 
scattering processes like $q t^c \rightarrow N \ell$.

First, as a warm-up, let us consider the case of
inverse decays $H \ell \rightarrow N$ in the 
single flavour approximation. 
It is well-known that
there is a CP asymmetry here, and this will show how
to apply CPT and unitarity bounds  
with unstable particles ($N$) in the final state.

Suppose that
\beq
|{\cal M}(H \ell \rightarrow N)|^2 \equiv \frac{|{\cal M}_0|^2}{2} 
(1 + \epsilon)\, , ~~~~
|{\cal M}(\bar{H} \bar{\ell} \rightarrow N)|^2 \equiv \frac{|{\cal M}_0|^2}{2} 
(1 - \epsilon)\, ,
\label{dec}
\eeq 
where $|{\cal M}_0|^2$ is order $\lambda^2$ and 
$|{\cal M}_0|^2  \epsilon$ is order $\lambda^4$. 

At  order $\lambda^2$, the CPT + unitarity
constraint is verified, but  at  order $\lambda^4$,
$|{\cal M}(H \ell \rightarrow N)|^2 \neq 
|{\cal M}(\bar{H} \bar{\ell} \rightarrow N)|^2$, 
 because there can be additional final states:
 $\bar{H} \bar{\ell}$  and $H \ell$. 
 As in the case of the Boltzman Equations, when
we include these $2 \leftrightarrow 2$ processes, we must
subtract out the real intermediate $N_1$, because
we have already included this by treating the $N_1$
as a final state particle. So at order $\lambda^4$:
\bea
|{\cal M}(H \ell \rightarrow X)|^2  & = & 
|{\cal M}(H \ell \rightarrow N)|^2 + 
|{\cal M}(H \ell \rightarrow \bar{H} \bar{\ell} )|^2 - 
|{\cal M}^{\rm RIS}(H \ell \rightarrow \bar{H} \bar{\ell} )|^2 
\nonumber \\
&+ & 
|{\cal M}(H \ell \rightarrow H \ell )|^2 - 
|{\cal M}^{\rm RIS}(H \ell \rightarrow H \ell )|^2
\nonumber \\
& = & 
 \frac{|{\cal M}_0|^2}{2} 
(1 + \epsilon) + |{\cal M}(H \ell \rightarrow \bar{H} \bar{\ell} )|^2
 - \frac{|{\cal M}_0|^2}{2} 
(1 + \epsilon)\frac{(1 + \epsilon)}{2}
\nonumber \\ &+&
 |{\cal M}(H \ell \rightarrow H \ell )|^2
 - \frac{|{\cal M}_0|^2}{2} 
(1 + \epsilon)\frac{(1 - \epsilon)}{2}  \nonumber \\
& = & 
 |{\cal M}(H \ell \rightarrow \bar{H} \bar{\ell} )|^2 +
  |{\cal M}(H \ell \rightarrow H \ell )|^2 +\cdots
\eea
where $X$ is all possible final states, and  the on-shell intermediate
$N_1$ has been subtracted in the narrow
width approximation  \cite{lept}. 
Repeating for $(\bar{H} \bar{\ell} \rightarrow X)$ will
give the same rate,  because  at ${\cal O}( \lambda^4)$,
$ |{\cal M}(H \ell \rightarrow \bar{H} \bar{\ell} )|^2$    
$ =  |{\cal M}(\bar{H} \bar{\ell} \rightarrow H \ell  )|^2$,  and
 $|{\cal M}(H \ell \rightarrow H \ell )|^2 = $
$ |{\cal M}( \bar{H} \bar{\ell} \rightarrow \bar{H} \bar{\ell} )|^2$.
So CPT and unitarity are satisfied for inverse decays,
as they ought to be.

CPT and unitarity are realised in the  scattering process
$(q t^c \rightarrow N \ell_\alpha)$, in a similar
way to  inverse decays. CPT and unitarity should hold order
by order in perturbation theory, so  we
work at order $\lambda^2 \lambda^2_\alpha h_t^2$, and define
\bea
|{\cal M}(q t^c \rightarrow N \ell_\alpha)|^2  & = & 
|{\cal M}_s|^2 (1 + \epsilon^{\alpha \alpha})\, , 
\label{scat}
\eea
where $|{\cal M}_s|^2 \propto \lambda^2_\alpha h_t^2$,
and $|{\cal M}_s|^2 \epsilon \propto \lambda^2 \lambda^2_\alpha h_t^2$. 
At order $ \lambda^2  \lambda^2_\alpha h_t^2$,  we should also include
various tree diagrams without $N$ in the final state.
Following the inverse decay discussion, one can write
\bea
|{\cal M}(q t^c \rightarrow X \ell_\alpha)|^2  & = & 
|{\cal M}(q t^c \rightarrow N \ell_\alpha)|^2 + 
\sum_\beta \left[ |{\cal M} (q t^c \rightarrow \ell_\beta H \ell_\alpha)|^2
-|{\cal M}^{\rm RIS} (q t^c \rightarrow \ell_\beta H \ell_\alpha)|^2  \right]
\nonumber \\
& + &   \sum_\beta \left[ |{\cal M} (q t^c \rightarrow 
\overline{\ell}_\beta \overline{H} \ell_\alpha)|^2
-|{\cal M}^{\rm RIS} (q t^c \rightarrow \overline{\ell}_\beta 
\overline{H} \ell_\alpha)|^2  \right]
\nonumber \\
& = & 
 |{\cal M}_s|^2 
(1 + \epsilon^{\alpha \alpha}) + 
|{\cal M}(q t^c \rightarrow  \ell H \ell_\alpha)|^2
 - |{\cal M}_s|^2 
(1 + \epsilon^{\alpha \alpha})\frac{(1 + \epsilon)}{2} \nonumber \\
&+&
  |{\cal M}(q t^c \rightarrow   \overline{\ell}  \overline{H} \ell_\alpha)|^2
 - |{\cal M}_s|^2 
(1 + \epsilon^{\alpha \alpha})\frac{(1 - \epsilon)}{2} \nonumber \\
& = & \sum_\beta \left[ |{\cal M}(q t^c \rightarrow \ell_\beta H \ell_\alpha)|^2
+  |{\cal M}(q t^c \rightarrow \overline{\ell}_\beta 
\overline{H} \ell_\alpha)|^2\right]\, ,
\label{ttNl}
\eea
where, in the narrow width approximation,
\beq
|{\cal M}^{\rm RIS} (q t^c \rightarrow \overline{\ell}_\beta \bar{H} \ell_\alpha)|^2 = 
|{\cal M}(q t^c \rightarrow N \ell_\alpha)|^2 \times 
{\rm BR} (  N \rightarrow \bar{H} \overline{\ell}_\beta)
\eeq

In eq.  (\ref{ttNl}), the CP asymmetry $\epsilon^{\alpha \alpha}$
has disappeared, so if  we repeat the calculation
for the CP conjugate initial state $\bar{q} \overline{t^c}$,
we should obtain the same result, verifying that 
a CP asymmetry in  $q t^c \rightarrow N \ell_\alpha$
 is consistent with CPT and unitarity. Furthermore, eq.
(\ref{ttNl}) is reassuring, because the unstable state $N$
has disappeared. There is no CP violation in the total rate
for $q t^c \rightarrow$ asymptotic  (stable) final states,
but CP violation in the partial rate to the unstable
$N$ is possible. This can be relevant to the
final value of the baryon asymmetry   when some of
the lepton flavours are weakly washed out (see section
\ref{4.1.2}).

\section{Approximate formulae for the baryon asymmetry }

In this section we present analytical formulae for the final baryon asymmetry
in the case in which flavours are taken into account.
We can have   different possible cases according to which interaction 
mediated by the charged
Yukawa couplings is in equilibrium, \cite{barbieri,nardietal}.

\subsection{$\mu$ and $\tau$  Yukawa couplings    
in equilibrium: $M_1\lsim 10^9$ GeV}

The $\mu$ and $\tau$ doublet leptons (and
by default, the electron doublet)
 will be mass eigenstates at the temperature of 
leptogenesis when the mass of the lightest right-handed  neutrino
$M_1$ is smaller  than about $10^9$ GeV. 
The  Boltzmann
equation for the diagonal entry $Y_{\alpha\alpha}$ reads
(no summation over the index $\alpha$)
\begin{equation}
\label{newdiagonal}
Y^\prime_{\alpha\alpha}=\epsilon_{\alpha\alpha}K z\, \frac{K_1(z)}
{K_2(z)}f_1(z)\Delta_{N_1}-\frac{1}{4}z^3 K_1(z)\, f_{2}(z)\,
K_{\alpha\alpha}Y_{\alpha\alpha}\, .
\end{equation}
Let us now solve analytically Eq. (\ref{newdiagonal}) according
to the magnitude of the various $K_{\alpha\alpha}$.
%
%\vskip 0.2cm
%\centerline
\subsubsection{ Strong wash-out regime for all flavours}
%\vskip 0.2cm
%\noindent

In such a case all the $K_{\alpha\alpha}\gg 1$. The right-handed neutrinos $N_1$'s are nearly in thermal equilibrium. Under these
circumstances, one can set $\Delta^\prime_{N_1}\simeq 0$ and
$\Delta_{N_1}\simeq (z K_2/4 g_* K)$. 
The lepton asymmetry for the flavour $\alpha$ is given by

\begin{equation}
Y_{\alpha\alpha}\simeq \epsilon_{\alpha\alpha}\int_0^{\infty}\, dz
\, \frac{K_1}{4 g_*}\,z^2\, e^{-\int_z^\infty\,dz^\prime
\, ((z^\prime)^3/4)K_1(z^\prime) K_{\alpha\alpha}}\, .
\end{equation}
Using the steepest descent method to evaluate the integral, one finds that
it gets the major contribution at $\overline{z}$ such that
$\overline{z}=\log\,K_{\alpha\alpha}+(5\ln\,\overline{z}/2)$ when
inverse decays become inefficient. The lepton asymmetry in the flavour
$\alpha$ becomes

\begin{equation}
\label{ll}
Y_{\alpha\alpha}\simeq 0.3\, \frac{\epsilon_{\alpha\alpha}}{g_*}
\left(\frac{0.55\times
10^{-3}\,{\rm eV}}{\tilde{m}_{\alpha\alpha}}
\right)^{1.16}\, .
\end{equation}
To get convinced that this result differs from the one usually considered 
in the literature, let us take 
the  total lepton asymmetry $Y_{\cal L}=\sum_\alpha 
Y_{\alpha\alpha}$

\begin{equation}
\label{o}
Y_{\cal L}\simeq \sum_\alpha 0.3\, \frac{\epsilon_{\alpha\alpha}}{g_*}
\left(\frac{0.55\times
10^{-3}\,{\rm eV}}{\tilde{m}_{\alpha\alpha}}
\right)^{1.16}\, .
\end{equation}
It does not coincide with total lepton asymmetry result (\ref{strongold})
in the strong wash-out regime. Indeed,
the total lepton asymmetry (\ref{o}) is the sum of the
$\epsilon_{\alpha\alpha}$, each weighted by the wash-out factor
$K_{\alpha\alpha}$
and not  the sum of the $\epsilon_{\alpha\alpha}$ divided
by the sum of the $K_{\alpha\alpha}$. Eqs. (\ref{strongold}) and
(\ref{o}) coincide only if one family is dominating the contribution
to the total CP asymmetry and the corresponding wash-out factor
is the tiniest.

\subsubsection{ Weak wash-out regime for all flavours}
\label{4.1.2}

In this case all the $K_{\alpha\alpha}\ll 1$. We assume that right-handed
neutrinos are not initially present in the plasma, but  they are generated
by inverse decays and scatterings. 
As explained in Sec. 2, the equation of motion for $Y_{N_1}$ is
well approximated by 

\begin{equation}
\label{1n}
Y_{N_1}'=-(K_s+K z)\left(Y_{N_1}-Y_{N_1}^{\rm EQ}\right)\, ,
\end{equation}
We split the solution in two pieces. Let us define $z_{\rm EQ}$ the value of 
$z$ at which $Y_{N_1}(z_{\rm EQ})=Y_{N_1}^{\rm EQ}(z_{\rm EQ})$. 
This value has to be found a posteriori. 
For $z\ll z_{\rm EQ}$, we may suppose that $Y_{N_1}\ll Y_{N_1}^{\rm EQ}$ and
Eq. (\ref{1n}) is solved by

\begin{eqnarray}
\label{2n}
Y^-_{N_1}(z)\simeq \int_0^z dz'\,(K_s+K z')Y_{N_1}^{\rm EQ}&=&
\frac{1}{4 g_*}\int_0^z dz'\,(K_s+K z')(z')^2\, K_2(z')\,  \nonumber \\
&=&\frac{K}{4 g_{\star}}\left( \frac{K_{s}}{K} I_{1}(z)+I_{2}(z) \right)
\end{eqnarray}
With $I_{1}$ and $I_{2}$ integral involving the modified Bessel functions :
\begin{equation}
I_{1}(z)=\int_{0}^{z} x^{2} K_{2}(x) dx \simeq f(z)+z^{3} K_{2}(z)
\end{equation}
where \cite{ogen} 
\begin{equation}
f(z)=\frac{3 \pi z^{3}}{\left( (9 \pi)^{c}+(2 z^{3})^{c} \right)^{1/c}},  c=0.7 
\end{equation} \\
The integral $I_{2}$ is well known, and equals
\begin{equation}
I_{2}(z)=\int_{0}^{z} x^{3} K_{2}(x) dx =8-z^{3}K_{3}(z)
\end{equation}
Therefore 
\begin{equation}
Y_{N_{1}}^{-}(z)\simeq \frac{K}{4 g_{\star}}\left(\frac{K_{s}}{K}(f(z)+z^{3}K_{2}(z))+8-z^{3}K_{3}(z) \right)
\end{equation}

As expected for weak washout, we find that the maximum
number density of $N_1$ is proportional to  $K$ (recall
$K_s \propto K$).

Let us now compute the value of $z_{\rm EQ}$. We expect it to be $\gg 1$ and 
we therefore approximate, up to ${\cal O} (z^{-3/2})$ : 
$K_{2}(z)\simeq K_{3}(z)\simeq \sqrt{\frac{\pi}{2}} z^{-1/2}e^{-z}$. Imposing 
$Y^-_{N_1}(z_{\rm EQ})=Y_{N_1}^{\rm EQ}(z_{\rm EQ})$, we find 
\begin{equation}
z_{\rm EQ}\simeq \frac{3}{2}\ln z_{\rm EQ} - \ln\left( \frac{8}{\sqrt{\pi /2}} K +3 \sqrt{\frac{\pi}{2}} K_{s}\right) \, .
\end{equation}
This solution is a good approximation to the real value for  $K\ll 1$. 

For $z>z_{\rm EQ}$, we have $(K_s+K z)\simeq K z$ and 

\begin{equation}
Y_{N_1}(z)\simeq Y_{N_1}^{\rm EQ}(z_{\rm EQ}) 
e^{K/2\left(z_{\rm EQ}^2-z^2\right)}\, .
\end{equation}

We have included  CP violation in $\Delta L = 1$ scattering,
unlike the usual analysis, so we expect our solution for
$Y^{\alpha \alpha}_L$ to have a different scaling with $K$ than 
eqn (\ref{weakold}).   The reason is as discussed in  \cite{lept,ogen}: 
if $C\!  P \! \! \! \! \!\! \! / ~~$ in scattering is neglected, then $Y_N \propto K$, and
the $N_1$ decay out of equilibrium, so one expects
$Y_L^{\alpha \alpha} \propto K \epsilon^{\alpha \alpha}$. 
However, if $C\!  P \! \! \! \! \!\! \! / ~~$  in $N_1$ production ($\simeq$ scattering)
is included, and washout is neglected, then  the
equations for $Y_{N_1}$ and $Y_{L}^{\alpha \alpha}$ are
identical, so  $Y_{L}^{\alpha \alpha} (z \rightarrow \infty)$ vanishes.
That is,  for every $|1/\epsilon^{\alpha \alpha}|$  $N_1$'s that 
are created, be it by inverse
decay or scattering, an (anti)-lepton $\alpha$  is produced. This
(anti-)asymmetry will approximately cancel against  
the lepton asymmetry generated later on, when
the $N_1$ decay.  However the cancellation will be imperfect,
because the anti-asymmetry has more time to
be washed out, so the final asymmetry should
scale as $K K_{\alpha \alpha}$.
After integrating by parts,
this is what we find for  the 
asymmetry in the flavour $\alpha$, 
which is  given by

\begin{eqnarray}
\label{z}
Y_{\alpha\alpha}&\simeq & \epsilon_{\alpha\alpha}\int_0^\infty\,dz' Y_{N_1}(z')
g_{\alpha\alpha}(z') e^{-\int_z^\infty dz'' g_{\alpha\alpha}(z'')}\, ,
\,\,\,
g_{\alpha\alpha}(z)=\frac{1}{4} z^3 K_{\alpha\alpha} K_1(z) f_2(z)\, ,
\nonumber\\
&\simeq& 
1.5\,  \frac{\epsilon_{\alpha\alpha}}{g_*}
\left(\frac{\tilde{m}_1}{3.3\times
10^{-3}\,{\rm eV}}\right)
\left(\frac{\tilde{m}_{\alpha\alpha}}{3.3\times
10^{-3}\,{\rm eV}}\right)\, .
\end{eqnarray}
We have checked numerically that this analytical formula fits  the 
numerical results to a 30\%.
 
Our findings hold provided that the non-resonant $\Delta L=2$
scattering rates, in particular those
mediated by the $N_2$ and $N_3$ heavy neutrinos, 
 are slower than decays and $\Delta L=1$ scatterings
when most of the asymmetry is generated. We 
estimate that this applies when 

\begin{equation}
\label{delta2}
\left(\frac{M_1}{10^{14}\, {\rm GeV}}\right)\ll 
 10^{-1}\, K_{\alpha\alpha}\, .
\end{equation}
This means that $K_{\alpha\alpha}$ should be larger than  $10^{-4}$.

\subsubsection{Strong wash-out  for some flavours and either 
weak or mild wash-out
for others}

In the  case in which $K_{\alpha\alpha}\gg 1$ for some flavour $\alpha$, but 
$K_{\beta\beta}\ll 1$ for some other flavour $\beta$,  it is
impossible to match this case with any of the cases  discussed 
in the section for the `one-flavour' approximation. 
Because some flavours $\alpha$ strongly interact with
the right-handed neutrinos, we may set $K=\sum_\alpha K_{\alpha\alpha}\gg 1$.
Thus, right-handed neutrinos are brought to thermal equilibrium
by inverse decays and by $\Delta L=1$ scatterings to an abundance
approximately given by 
%\Delta^\prime_{N_1}\simeq 0$ and
%$\Delta_{N_1}\simeq (z K_2/4 g_* K f_1)$.

\begin{equation}
Y_{N_1}(z)\simeq Y_{N_1}^{\rm EQ}\left(1-e^{-K_s z-K z^2/2}\right)\, .
\end{equation}
The lepton asymmetry in  flavour $\beta$ is then given by

\begin{eqnarray}
\label{third}
Y_{\beta\beta}&\simeq& \epsilon_{\beta\beta}\,
\int_{0}^{\infty}\,dz\, 
%Y_{N_1}^{\rm EQ}(z')\left(1+\frac{1}{f_(z')}\right)
Y_{N_1}^{\rm EQ}(z')
\left(1-e^{-K_s z'-K z^{'2}/2}\right)
g_{\beta\beta}(z')\nonumber\\
&\simeq& 0.4\, \frac{\epsilon_{\beta\beta}}{g_*}
\left(\frac{\tilde{m}_{\beta\beta}}{3.3\times
10^{-3}\,{\rm eV}}\right)\, .
\end{eqnarray}
Again, we have checked that this formula  fits the numerical results
to about 30 \%. 
It has the dependence on  $K_{\beta \beta}$ expected
from our inclusion of  $C\!  P \! \! \! \! \!\! \! / ~~$ in scattering.
The anti-asymmetry $ \sim -\epsilon_{\beta \beta} s/g_*$ created
during $N_1$ production has the leisure to be partially destroyed by processes
violating $\beta$ lepton number.  The amount of this reduction
is controlled by $K_{\beta \beta}$, so one expects the final $\beta$
asymmetry  $ \propto  \epsilon_{\beta \beta} K_{\beta \beta}$,
as in eq. (\ref{third}). Again, this result applies for $K_{\alpha\alpha}
\gg 10^{-4}$. 

In the  case in which $K_{\alpha\alpha}\gg 1$ for some flavour $\alpha$ and 
the 
flavour $\beta$ suffers a  mild ($K_{\beta\beta}\sim 1$) 
wash-out, the final asymmetry in the flavour
$\beta$ is fitted within 30 \% by the formula

\begin{equation}
\label{interpolate}
Y_{\beta\beta}\simeq \frac{\epsilon_{\beta\beta}}{g_*}
\left(
\left(\frac{\tilde{m}_{\beta\beta}}{8.25\times
10^{-3}\,{\rm eV}}\right)^{-1}+
\left(\frac{0.2\times
10^{-3}\,{\rm eV}}{\tilde{m}_{\beta\beta}}
\right)^{-1.16}\
\right)^{-1}\, .
\end{equation}
This formula reduces to Eqs. (\ref{ll}) and (\ref{third}) for the
weak and strong wash out for the flavour $\beta$,  respectively.

\subsubsection{Recipe to go from the flavour asymmetries to the
baryon asymmetry}
\label{recipe1}

The final baryon asymmetry is given by \cite{nardietal}\footnote{if
the sphalerons freeze out before the eletroweak phase transition,
12/37 would be replaced by another fraction\cite{mikko} of order 1/3, 
such as\cite{Harvey:1990qw} 28/79.}

\begin{equation}
Y_{\cal B}\simeq \frac{12}{37}\, \sum_\alpha Y_{\Delta_\alpha}\, ,
\end{equation}
where  the asymmetries $Y_{\Delta_\alpha}$, 
($\Delta_{\alpha}=( B/3 - L_{\alpha})$)
are  conserved by the sphaleron  and other Standard Model 
interactions. To provide an analytical 
formulae for the baryon asymmetry,
in the case in which $M_1\lsim 10^9$ GeV, it suffices 
to  obtain the  asymmetries $Y_{\Delta_\alpha}$ from
our formulae for 
the lepton flavour  asymmetries  $Y_{\alpha\alpha}$.

In this paper, we approximate $Y_{\Delta_\alpha} = 
Y_{\alpha\alpha}( \tilde{m}_{\alpha \alpha} 
\rightarrow A_{\alpha \alpha} \tilde{m}_{\alpha \alpha})$.
That is, in our approximate solutions  for $Y_{\alpha \alpha}$,
we replace $\tilde{m}_{\alpha \alpha}$ by  
$A_{\alpha \alpha} \tilde{m}_{\alpha \alpha}$ (no sum on
$\alpha$),  where the matrix 
$A$ \cite{barbieri} is given in eqns (\ref{ATmu}) and (\ref{ATtau}).
This approximation should be adequate for transforming the flavour
asymmetries, estimated flavour by flavour, 
into a baryon asymmetry.  To see this, we review the
discussion of \cite{barbieri} :

The interactions which are much faster than $H$, such as
gauge interactions, sphalerons, and some Yukawas, can
be approximately taken into account by imposing chemical 
equilibrium conditions(see \cite{nardietal1, nardietal}
for a  more careful analysis). This allows to express the asymmetries
in all particle species as linear combinations of
the asymmetries in conserved quantum numbers.  These  interactions-in-equilibrium respect the three flavoured asymmetries
$\Delta_\alpha = B/3 - L_\alpha$, which are changed by the interactions of $N$. So
it is the Boltzmann Equation for $ Y_{\Delta_\alpha}$
that is relevant for leptogenesis, and it  is clearly \cite{barbieri} 
 the right-hand-side
of eqn (\ref{new2a}):
 $$\frac{d Y_{\Delta_\alpha}}{dz} =
    \frac{z}{s H(M_1)}\left[\left(
\frac{Y_{N_{1}}}{Y_{N_{1}}^{\rm EQ}} - 1\right)
\epsilon^{\alpha \alpha} (\gamma_D +\gamma_{\Delta L =1}) -
\frac{Y^{\alpha \alpha}}{Y_L^{\rm EQ}} \left(\gamma_D^{\alpha \alpha}
+\gamma_{\Delta L =1}^{\alpha \alpha} %+ \gamma_{\Delta L=2}
\right)\right]\, .
% {\rm ~ the ~right~hand~side~ 
%of~ eqn~ (\ref{new2a})},
$$ 
 $Y_{\alpha \alpha}$ on the right-hand-side can be
re-expressed as $Y_{\alpha \alpha} = \sum_\beta A_{\alpha \beta}
 Y_{\Delta_\beta}$.  An
$\ell_\alpha$  produced or destroyed in the plasma changes
$B/3 - L_\alpha$, but not all the $B/3 - L_\alpha$ is in
 $Y_{\alpha \alpha}$, so washout is reduced and 
the flavoured Boltzmann Equations are coupled.

In the temperature range where the $\mu$ Yukawa is in equilibrium,
the $A$ matrix is given as \footnote{This disagrees with
our previous version\cite{davidsonetal}, where we had
 taken the $u$ and $d$ Yukawas in equilibrium,
and differs from  \cite{barbieri} in that we  include
strong sphalerons. Our  results agree with \cite{nardietal1,nardietal}.
}
\begin{equation}
\left(\begin{array}{c}
Y_{ee} \\
Y_{\mu\mu} \\
Y_{\tau\tau}
\end{array}\right) =  \left(\begin{array}{ccc}
-151/179 & 20/179 & 20/179\\
25/358 & -344/537 & 14/537\\
25/358 &  14/537 &  -344/537
\end{array}\right)\left( \begin{array}{c}
Y_{\Delta_e} \\Y_{\Delta_{\mu}} \\ Y_{\Delta_{\tau}}
\end{array}\right)\, .
\label{ATmu}
\end{equation}
The off-diagonal elements of $A$ couple the flavoured Boltzmann
Equations.  We neglect this effect, estimating 
that it has little effect on the final baryon asymmetry:
since the off-diagonals of $A$ are small,
$A_{\alpha \beta} Y_{\Delta_\beta}$ (for $\beta \neq \alpha$,
not summed) will only make a significant contribution
to the washout of  $Y_{\Delta_\alpha}$ when  
 $Y_{\Delta_\beta} \gg Y_{\Delta_\alpha}$.  In this
case, the contribution of  $Y_{\Delta_\alpha}$ to
the baryon asymmetry is small, so  it is
of little numerical importance that our formula mis-estimated
the washout in flavour $\alpha$.

Thus the recipe to compute the baryon asymmetry in the case
in which $M_1\lsim 10^9$ GeV is the following:
\noindent

1) Compute the $K_{\alpha\alpha}$ parameters for the three flavours
to see  which of the three different previously discussed cases applies;
\noindent

2) compute the  CP asymmetry $\epsilon_{\alpha \alpha}$
 in each flavour, and approximate the  asymmetry 
$Y_{\Delta_\alpha}$ using the relevant
formualae for $Y_{\alpha \alpha}$ 
among equations (\ref{o}), (\ref{z}),  (\ref{third}),
{\it but with $\tilde{m}_{\alpha \alpha}$ replaced by 
 $|A_{\alpha \alpha}| \tilde{m}_{\alpha \alpha}$ } (no sum on $\alpha$). 
\noindent

3) compute the baryon asymmetry

\begin{equation}
Y_{\cal B}\simeq - \frac{12}{37} 
\left[  Y_{ee}\left(\epsilon_{ee}, \frac{151}{179}\tilde{m}_{ee}\right)
+ Y_{\mu \mu}\left(\epsilon_{\mu \mu}, \frac{344}{537}
\tilde{m}_{\mu \mu}\right)
+ Y_{\tau \tau}\left(\epsilon_{\tau \tau}, \frac{344}{537}
\tilde{m}_{\tau \tau}\right) \right]
\label{ppYB}
\end{equation}

Figs. {\ref{strong}, {\ref{weak} and {\ref{mixed} illustrate the differences
between the final baryon asymmetry with flavours accounted for and the result obtained
within the one-flavour approximation for some values of the wash-out
parameters and CP asymmetries.

%%%%%%%%%%%%%%%%%%%%%%%%%%%%%%%%%%%%%%%%%%%%%%%%%%%%%%%%%%%
\begin{figure}[h]%%%%%%%%%%%%%%%%%%%%%%%%%%%%%%%%%%%%%%%%%%%
  %%%
  \centerline{\hspace{-1.5cm}
%    \scalebox{0.75}{\includegraphics{strong.eps}}
 \scalebox{0.45}{\includegraphics{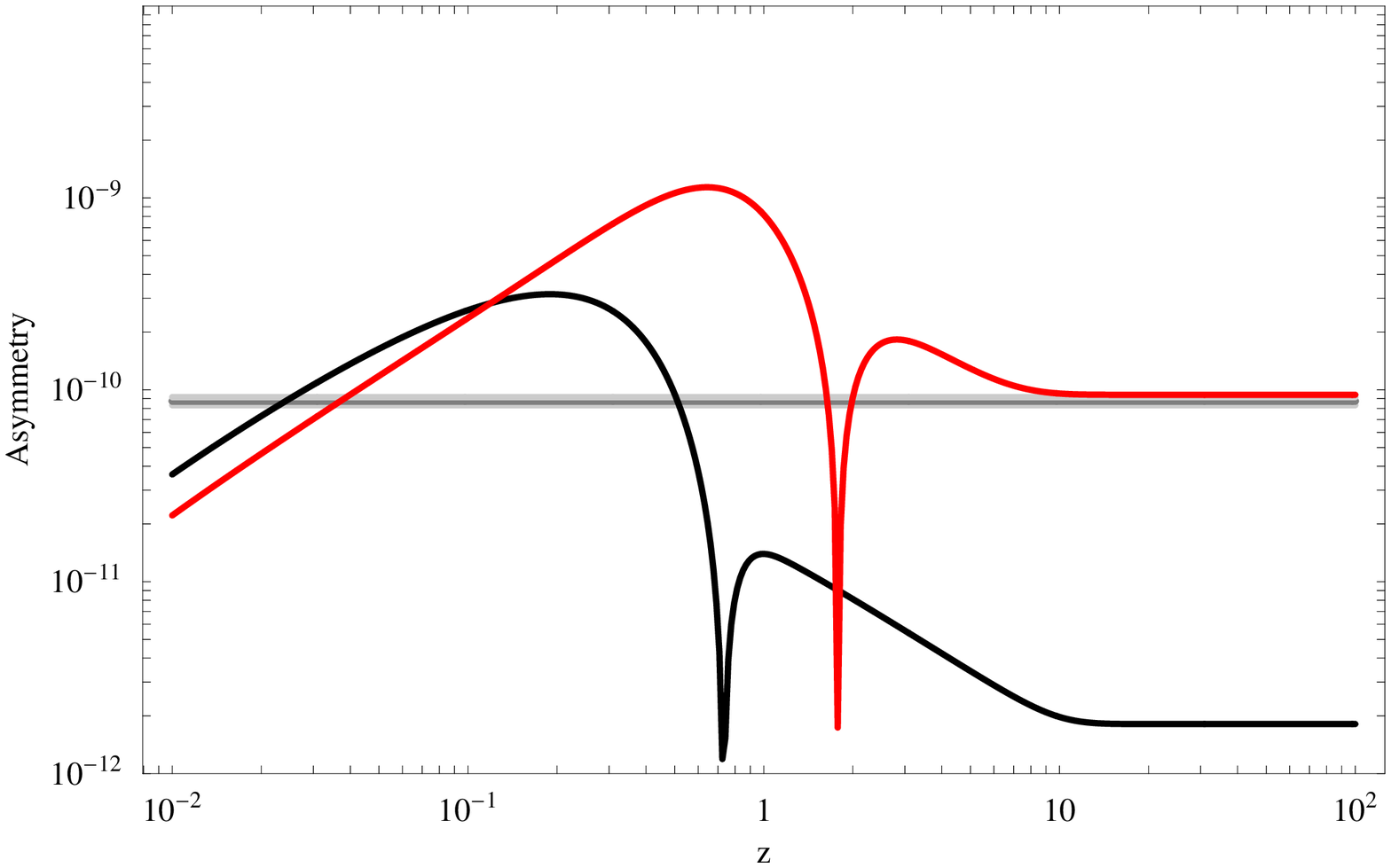}}}
   \caption{\small The total baryon asymmetry in the two flavour calculation
 (upper) 
and within the one-flavour approximation (lower) as a function of $z$. The
chosen parameters are $K_{\tau \tau}=10$, 
$K_{\mu\mu}=30$, $K_{ee}=30$, 
$\epsilon_{\tau \tau}=2.5\times 10^{-6}$, $\epsilon_{\mu\mu}=-2\times 10^{-6}$,
$\epsilon_{ee}=10^{-7}$ and $M_1=10^{10}$ GeV. {\it See the second
Note Added at the beginning of the manuscript.}
}
  \label{strong}
\end{figure}
\vspace{.5cm}

%%%%%%%%%%%%%%%%%%%%%%%%%%%%%%%%%%%%%%%%%%%%%%%%
%%%%%%%%%%%
%%%%%%%%%%%%%%%%%%%%%%%%%%%%%%%%%%%%%%%%%%%%%%%%%%%%%%%%%%%
\begin{figure}[h]%%%%%%%%%%%%%%%%%%%%%%%%%%%%%%%%%%%%%%%%%%%
  %%%
\vspace{.5cm}
  \centerline{\hspace{-1.5cm}
% \scalebox{0.75}{\includegraphics{weak.eps}}
    \scalebox{0.45}{\includegraphics{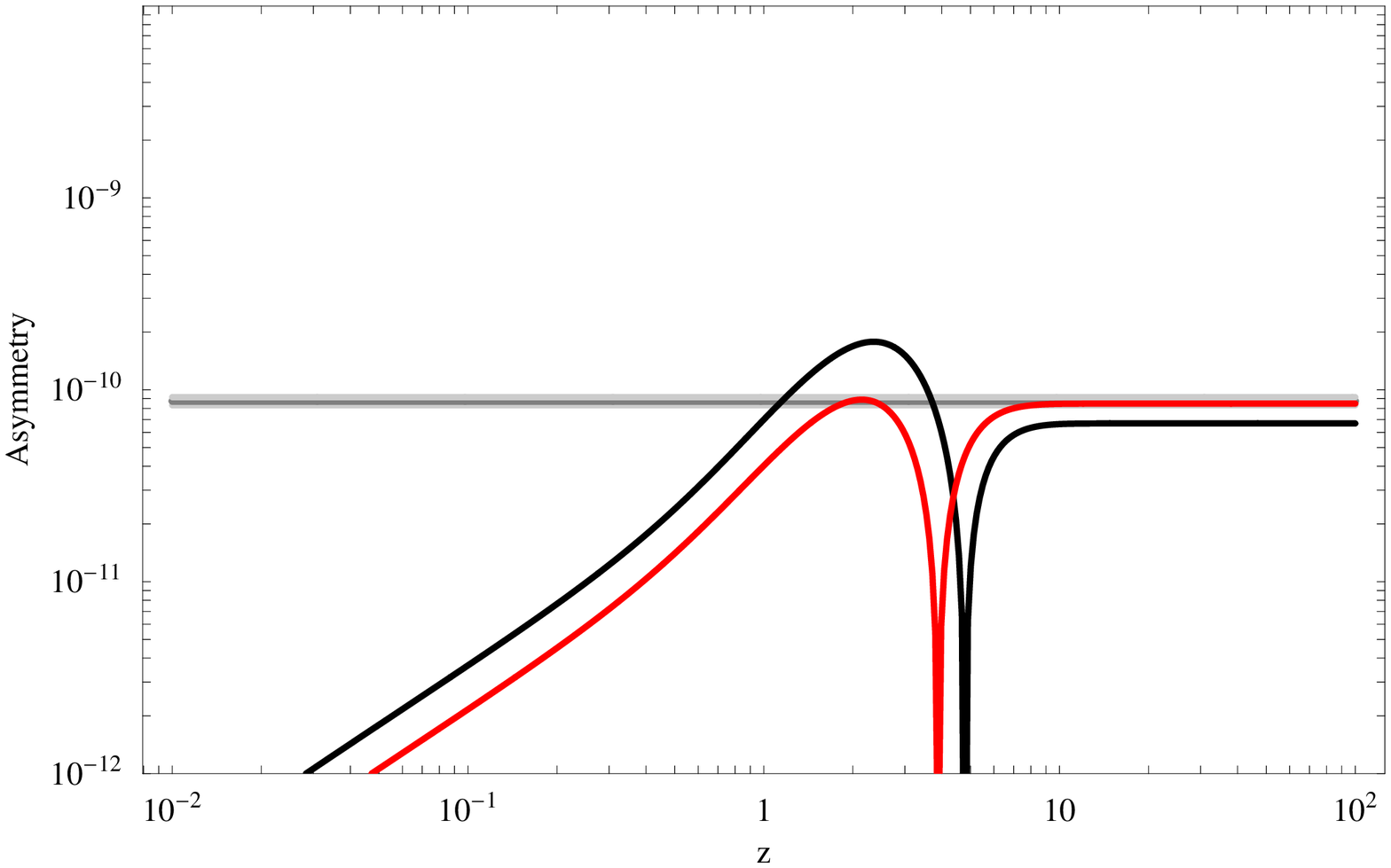}}}
   \caption{\small The total baryon asymmetry in the two flavour calculation (upper) 
and within the one-flavour approximation (lower) as a function of $z$. The
chosen parameters are $K_{\tau \tau}=4.5 \times 10^{-1}$, 
$K_{\mu\mu}=10^{-2}$, $K_{ee}=10^{-3}$, 
$\epsilon_{\tau \tau}=2.5\times 10^{-6}$, $\epsilon_{\mu\mu}=-2\times 10^{-6}$,
$\epsilon_{ee}=10^{-7}$ and $M_1=10^{10}$ GeV.  {\it See the second
Note Added at the beginning of the manuscript.}
}
  \label{weak}
\end{figure}%%%%%%%%%%%%%%%%%%%%%%%%%%%%%%%%%%%%%%%%%%%%%%%%
%%%%%%%%%%%
\begin{figure}[h]%%%%%%%%%%%%%%%%%%%%%%%%%%%%%%%%%%%%%%%%%%%
  %%%
  \centerline{\hspace{-1.5cm}
% \scalebox{0.75}{\includegraphics{mixed.eps}}
    \scalebox{0.45}{\includegraphics{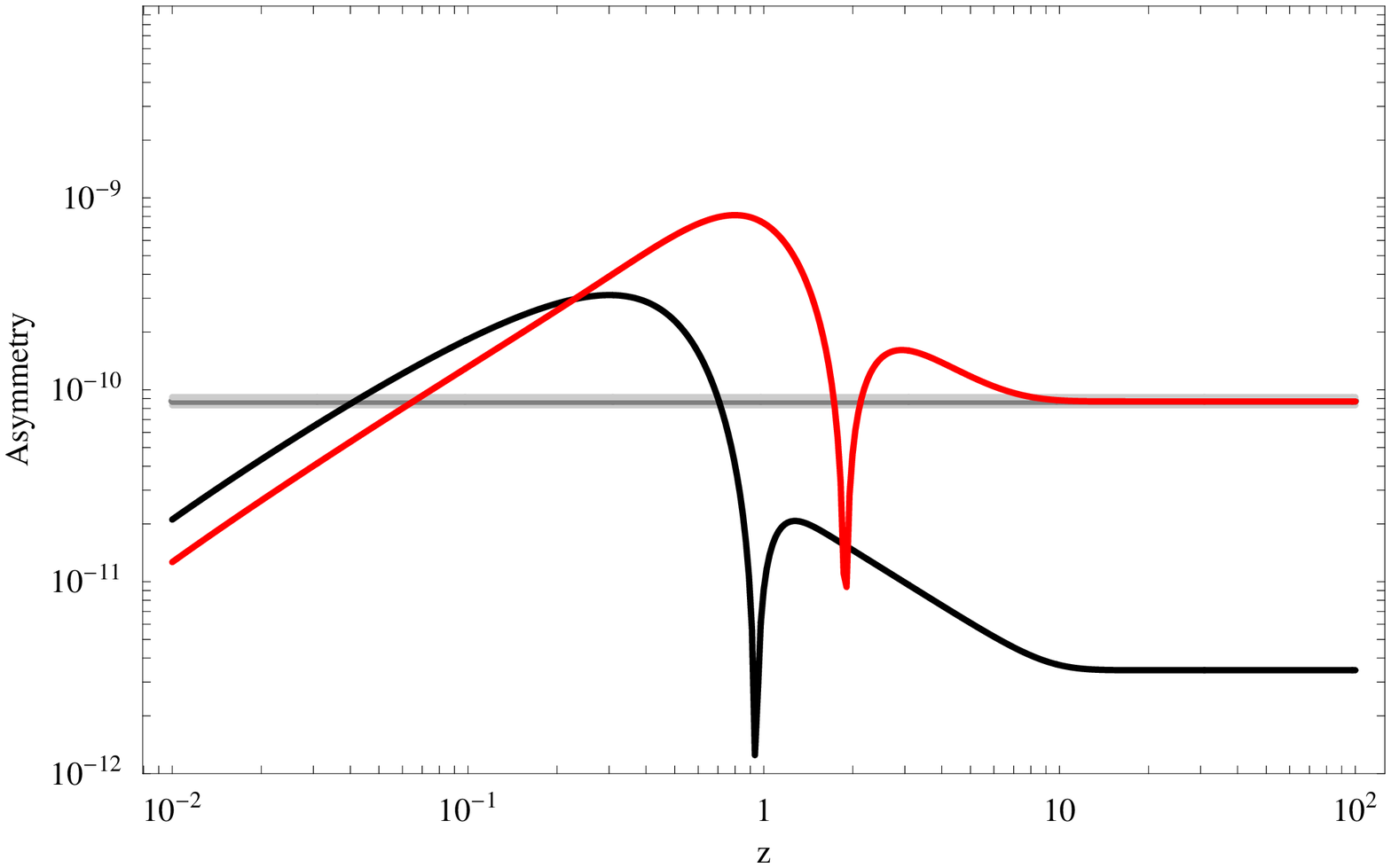}}}
   \caption{\small The total baryon asymmetry in the two flavour calculation (upper) 
and within the one-flavour approximation (lower) as a function of $z$. The
chosen parameters are $K_{\tau \tau}=10$, 
$K_{\mu\mu}=30$, $K_{ee}=10^{-2}$, 
$\epsilon_{\tau \tau}=2.5\times 10^{-6}$, $\epsilon_{\mu\mu}=-2\times 10^{-6}$,
$\epsilon_{ee}=10^{-7}$ and $M_1=10^{10}$ GeV.  {\it See the second
Note Added at the beginning of the manuscript.}
}
  \label{mixed}
\end{figure}%%%%%%%%%%%%%%%%%%%%%%%%%%%%%%%%%%%%%%%%%%%%%%%%
%%%%%%%%%%%

\subsection{Only the tau Yukawa coupling in equilibrium: $(10^9\lsim M_1\lsim
10^{12})$ GeV}
\label{tau}
This case is realized when the mass of the lightest right-handed  neutrino
$M_1$ is larger   than about $10^9$ GeV, but smaller than
about $10^{12}$ GeV\footnote{In the case in which $M_1$ is around $10^9$ GeV
off-diagonal terms may be relevant. However,  they are quickly 
damped away as soon as $M_1$ becomes larger than $10^9$ GeV 
\cite{davidsonetal}.}. Interactions mediated by $h_\tau$ are in equilibrium,
but not those mediated by $h_\mu$. 
The off-diagonal entries
of the matrix $Y_{\tau\beta}$ all vanish and the muon and electron
asymmetries are indistinguishable. The  
problem of finding the total baryon asymmetry 
reduces to a case of two flavours, the lepton $\ell_\tau$,
 and $\hat{\ell}_2$,  the non-$\tau$ components of the  lepton into
which $N_1$ decays.  At  tree level,
$\ell_2 =\sum_{\alpha=e,\mu}\lambda_{1\alpha}\ell_\alpha/
\left(\sum_{\alpha=e,\mu}\left|\lambda_{1\alpha}\right|^2\right)^{1/2}$. 
One can therefore define two CP asymmetries, $\epsilon_{\tau\tau}$ and
$\epsilon_2=\epsilon_{ee}+\epsilon_{\mu\mu}$,  and the corresponding
wash-out parameters $K_{\tau\tau}$ and $K_2=K_{ee}+K_{\mu\mu}$ for the
two asymmetries $Y_{\tau\tau}$ and $Y_{2}=Y_{ee}+Y_{\mu\mu}$. 
One then solves this set of equations as we described in the
previous subsection depending upon the magnitude of the wash-out parameters
$K_{\tau\tau}$ and $K_2$.

The recipe to compute the baryon asymmetry in the case
in which $(10^9\lsim M_1\lsim 10^{12})$ GeV is the following:
\noindent

1) Compute the $K_{\tau\tau}$ and $K_2$ parameters 
to see  whether they are either both larger than unity, or both smaller than
unity, or one larger and the other smaller than unity;
\noindent

2) compute the CP asymmetry in each of the two flavours, and
the $B/3 - L_\alpha$  asymmetry $Y_{\Delta_\alpha}$ using the relevant
formulae among equations (\ref{o}), (\ref{z}),  (\ref{third}),
{\it replacing in these formulae $\tilde{m}_{\beta \beta}$ by
$|A_{\beta \beta}| \tilde{m}_{\beta \beta}$} (no sum on $\beta$), where
in this case the $A$ matrix\cite{barbieri} is \footnote{This differs
slightly from results in  \cite{barbieri, nardietal} because
we consider the case where the  $c$
Yukawa and strong sphalerons are in equilibrium}
%\begin{equation}
%\left(\begin{array}{c}
%Y_{ee} \\
%Y_{\mu\mu} \\
%Y_{\tau\tau}
%\end{array}\right) =  \left(\begin{array}{ccc}
%-317/351 & 34/351 & 20/351\\
% 34/351& -317/351  & 20/351\\
% 1/117 & 1/117 & -82/117
%\end{array}\right)\left( \begin{array}{c}
%Y_{\Delta_e} \\Y_{\Delta_{\mu}} \\ Y_{\Delta_{\tau}}
%\end{array}\right)\, .
%\label{ATtau}
%\end{equation}

\begin{equation}
\left(\begin{array}{c}
Y_{2} \\
Y_{\tau\tau}
\end{array}\right) =  \left(\begin{array}{cc}
  -417/589  & 120/589\\
   30/589 & -390/589
\end{array}\right)\left( \begin{array}{c}
Y_{\Delta_{2}} \\ Y_{\Delta_{\tau}}
\end{array}\right)\, .
\label{ATtau}
\end{equation}

\noindent

3) compute the baryon asymmetry

\begin{equation}
Y_{\cal B}\approx -\frac{12}{37}\left( Y_{2}(\epsilon_{2}, 
417\tilde{m}_{2}/589)+ Y_{\tau \tau}
(\epsilon_{\tau \tau}, 390\tilde{m}_{\tau \tau}/589)
\right)\, .
\label{BAUM1large}
\end{equation}
 In  this case,  the first row, second colomn
element of the matrix (\ref{ATtau}) is $\sim 30 \%$ of the
diagonal elements, so the approximation of
neglecting off-diagonals is less good than in the $T \lsim 10^9$ GeV case.
Again, our results apply if  $\Delta L=2$ scatterings are negligible, see
eq. (\ref{delta2}).

\subsection{No charged Yukawa couplings in equilibrum: $M_1\gsim 10^{12}$ GeV}
This case is realized when the mass of the lightest right-handed  neutrino
$M_1$ is larger   than about $10^{12}$ GeV.  Interactions mediated by 
all charged lepton Yukawa couplings are out of  equilibrium and all flavours are
indistinguishable. The  
problem of finding the total baryon asymmetry 
reduces to a case of one  flavour,  
the lepton $\hat{\ell}_1=\sum_{\alpha=e,\mu,\tau}\lambda_{1\alpha}\ell_\alpha/
\left(\sum_{\alpha=e,\mu,\tau}\left|\lambda_{1\alpha}\right|^2\right)^{1/2}$. 
One can therefore define a single  CP asymmetry, 
$\epsilon_1=\sum_\alpha\epsilon_{\alpha\alpha}$,  and the corresponding
wash out parameter $K=\sum_\alpha K_{\alpha\alpha}$. The corresponding 
Boltzmann equations read as those in Eq. (\ref{old1}). 
Only in this
case the commonly used formulae reported in  Section 2
hold with the eventual inclusion of the effects from the
$\Delta L=2$ scatterings.

\section{General analysis of leptogenesis in the 
two right-handed neutrino model}

In this section we will concentrate on the see-saw model with
two right-handed neutrinos (2RHN). Oscillation experiments
indicate that two new mass scales have to be introduced, in order 
to account for the solar and atmospheric mass splittings.
These two mass scales could be associated to the masses of 
two right-handed neutrinos, therefore a see-saw model with just two
right-handed neutrinos can already accommodate all the observations.
An additional motivation to consider the two right-handed neutrino
model is that it corresponds to some interesting limits
of the complete see-saw model with three right-handed
neutrinos, namely when the mass of the heaviest
right-handed neutrino is much larger than the masses of the
other two, or when the Yukawa couplings for the first
generation of right-handed neutrinos are much smaller than
for the other two generations.

The two right-handed neutrino model depends on many less parameters
than the complete see-saw model and the theoretical analysis
of leptogenesis becomes much more manageable, while preserving
the key features of the model with three right-handed neutrinos.
In the basis where the charged lepton Yukawa coupling and the
right-handed mass matrices are diagonal, the model is defined
at high energies by a $2\times 3$ Yukawa
matrix and two right-handed neutrino masses, $M_1$ and $M_2$.
This amounts to eight moduli and three phases. On the other hand, at low
energies the neutrino mass matrix is defined by five moduli
(two masses and three mixing angles) plus two phases (the Dirac phase
and the Majorana phase). In particular,
since the mass matrix is rank 2 in this model,
the lightest neutrino mass eigenvalue automatically vanishes
and only two possible spectra may arise:
\begin{itemize}
{\item Normal hierarchy: $m_1=0$ ,
$m_2=\sqrt{\Delta m^2_{sol}}$ , $m_3=\sqrt{\Delta m^2_{atm}}$ }
{\item Inverted hierarchy:  $m_3=0$, 
$m_1=\sqrt{\Delta m^2_{atm}-\Delta m^2_{sol}}$ , 
 $m_2=\sqrt{\Delta m^2_{atm}}$ }
\end{itemize}
The only Majorana phase corresponds to the 
phase difference between the two non-vanishing mass eigenvalues.
Therefore, the number of unmeasurable parameters in the
2RHN model is reduced to three moduli and one phase.

The most general Yukawa coupling
compatible with the low energy data is given by:
\begin{equation}
\lambda=M^{1/2} R m^{1/2} U^{\dagger }/v,   \label{yukawa}
\end{equation}
where  $m={\rm Diag}(m_1,m_2,m_3)$ is  
the diagonal matrix of the light neutrino mass eigenvalues
(which has $m_1=0$ for the  normal hierarchy
and $m_3=0$ for the inverted hierarchy),
$M={\rm Diag}(M_1,M_2)$ is the diagonal matrix of the right
handed neutrino masses, $U$ is the leptonic
mixing matrix, and $R$ is an orthogonal matrix that 
in the 2RHN model has the following structure 
\cite{Ibarra:2003xp}
\begin{equation}
R=\left( 
\begin{array}{ccc}
0 & \cos \hat\theta & \xi \sin \hat\theta \\ 
0 & -\sin \hat\theta & \xi \cos \hat\theta  
\end{array} \right) (\rm{normal~hierarchy}), 
\end{equation}
\begin{equation}
R=\left( 
\begin{array}{ccc}
\cos \hat\theta & \xi \sin \hat\theta & 0\\  
-\sin \hat\theta & \xi \cos \hat\theta  & 0
\end{array} \right) (\rm{inverted~hierarchy}),  \label{R2x3}
\end{equation}
with $\hat\theta$ a complex parameter
and $\xi=\pm 1$ a discrete parameter that accounts for a 
discrete indeterminacy in $R$. 
In consequence, the elements of the neutrino Yukawa matrix read:
\begin{eqnarray}
\lambda_{1\alpha} &=&\sqrt{M_{1}}(\sqrt{m_{2}}\cos \hat\theta~U_{\alpha2}^{\ast}+\xi
\sqrt{m_{3}}\sin \hat\theta~U_{\alpha3}^{\ast })/v,  \nonumber \\
\lambda_{2\alpha} &=&\sqrt{M_{2}}(-\sqrt{m_{2}}\sin \hat\theta~U_{\alpha2}^{\ast }+\xi 
\sqrt{m_{3}}\cos \hat\theta~U_{\alpha3}^{\ast })/v,  
\label{yukawa-elements-norm}
\end{eqnarray}
for the case with normal hierarchy and
\begin{eqnarray}
\lambda_{1\alpha} &=&\sqrt{M_{1}}(\sqrt{m_{1}}\cos \hat\theta~U_{\alpha1}^{\ast}+\xi
\sqrt{m_{2}}\sin \hat\theta~U_{\alpha2}^{\ast })/v, \nonumber  \\
\lambda_{2\alpha} &=&\sqrt{M_{2}}(-\sqrt{m_{1}}\sin \hat\theta~U_{\alpha1}^{\ast }+\xi 
\sqrt{m_{2}}\cos \hat\theta~U_{\alpha2}^{\ast })/v,  
\label{yukawa-elements-inv}
\end{eqnarray}
for the case with inverted hierarchy. The three moduli and the phase
that are not determined by low energy experiments are
identified in this parametrization with the two right-handed 
masses $M_1$ and $M_2$, and the complex parameter $\hat \theta$.

Notice that we have included all the low
energy phases in the definition of the matrix $U$, 
{\it i.e.} we have written the leptonic mixing matrix in the 
form $U=V~{\rm Diag}(1, e^{-i\phi/2},1) $,
where $\phi $ is  the Majorana phase and $V$ has
the form of the CKM matrix:
\bea
V=\pmatrix{c_{13}c_{12} & c_{13}s_{12} & s_{13}e^{-i\delta}\cr
-c_{23}s_{12}-s_{23}s_{13}c_{12}e^{i\delta} & c_{23}c_{12}-s_{23}s_{13}s_{12}e^{i\delta} & s_{23}c_{13}\cr
s_{23}s_{12}-c_{23}s_{13}c_{12}e^{i\delta} & -s_{23}c_{12}-c_{23}s_{13}s_{12}e^{i\delta} &
c_{23}c_{13}\cr},
\label{Vdef}  
\eea
so that the neutrino mass matrix is 
${\cal M}=U^* {\rm Diag}(m_1,m_2,m_3) U^{\dagger}$.
It is straightforward to check that the Yukawa coupling eq.(\ref{yukawa})
indeed satisfies the see-saw formula 
${\cal M}= \lambda ^T {\rm Diag}(M^{-1}_1,M^{-1}_2) \lambda v^2$.

The flavour CP asymmetries can be readily
computed in terms of low energy data and the unmeasurable
parameters $M_1$, $M_2$ and $\hat\theta$ substituting
the expression for the Yukawa coupling 
in eq.(\ref{flavour-CPasym}). In the limit 
$M_1\ll M_2$ we obtain for the case with normal hierarchy 
the following result:
\bea
\epsilon_{\alpha\alpha}&\simeq& \frac{3}{8\pi v^2}
\frac{M_1}{m_2|\cos^2\hat\theta|+m_3|\sin^2\hat\theta|}
[(m^2_3|U_{\alpha3}|^2-m^2_2|U_{\alpha2}|^2)\;{\rm Im} \sin^2\hat\theta + \nonumber\\
&&+\xi \sqrt{m_2 m_3}(m_3+m_2)\;{\rm Re}U^*_{\alpha2}U_{\alpha3}\; 
{\rm Im}\sin \hat\theta\cos \hat\theta+ \nonumber\\
&&+\xi \sqrt{m_2 m_3}(m_3-m_2)\;{\rm Im}U^*_{\alpha2}U_{\alpha3}
\;{\rm Re}\sin \hat\theta\cos \hat\theta],
\label{CP-flavour-2RHN}
\eea
and analogously for the case with inverted hierarchy, with the
changes in the labels $3\rightarrow 2$ and $2\rightarrow 1$.

We will analyze numerically the predictions for the baryon
asymmetry taking flavour properly into account,
%solving eqs.(\ref{new1b},\ref{new2b}), 
as described in Section 3, and also for comparison
following the conventional computation ignoring flavour, 
%solving  eqs(\ref{old2},\ref{old1}),
as described in 
Section 2. We will perform this analysis both
for the case of normal hierarchy and inverted 
hierarchy, fixing the atmospheric and solar mass splittings
and  mixing angles 
to the values suggested by oscillation experiments,
$\Delta m^2_{atm}\simeq 2.2\times 10^{-3}\eV^2$,
$\Delta m^2_{sol}\simeq 8.1\times 10^{-5}\eV^2$,
$\theta_{23}\simeq\pi/4$ and $\theta_{12}\simeq \pi/6$,
respectively \cite{Maltoni:2004ei}. The remaining parameters in the leptonic
mixing matrix are fixed to $\theta_{13}=0.1$,
$\delta=\pi/4$, $\phi=\pi/3$. In general, the results are 
not very sensitive to the value of $\theta_{13}$. On the other
hand, the aspect of the plots does depend on the precise value
of the phases $\delta$ and $\phi$, although our main 
conclusions remain valid.

We will show our results for different values of
$M_1$, to cover the possibility that only the tau Yukawa
interaction is in equilibrium ($M_1\gsim 10^9\GeV$) or
that the tau and the muon Yukawa interactions are in equilibrium
($M_1\lsim 10^9\GeV$), and for different values of the complex 
parameter $\hat\theta$,
restricting ourselves to the region $|\hat\theta|<1$. We will
also fix $\xi=1$, although the results for the case $\xi=-1$
can be read from our results by changing $\hat\theta\rightarrow -\hat\theta$,
as can be checked from eq.(\ref{CP-flavour-2RHN}). 

In Fig.\ref{normal} we show the result of the calculation
of the baryon asymmetry in the complex plane of $\hat\theta$ 
when the mass spectrum presents a normal hierarchy,
following the calculation that takes flavour
properly into account (left plots) and following the
conventional calculation ignoring flavour (right plots).
In the upper plots we show the results when $M_1=10^8\GeV$,
so that the tau and muon Yukawa interactions are in 
equilibrium, whereas in the lower plots we take $M_1=10^{10}\GeV$,
so that only the tau Yukawa interaction is in equilibrium. 
It is apparent from these plots that the proper treatment of flavour 
in the Boltzmann equations is necessary in order to calculate correctly the
baryon asymmetry.\footnote{In particular, the prediction for the
baryon asymmetry in the conventional computation is symmetric
under ${\rm Re}\hat\theta\rightarrow -{\rm Re}\hat\theta$
and ${\rm Im}\hat\theta\rightarrow -{\rm Im}\hat\theta$,
and consequently independent of the discrete parameter $\xi$.
On the other hand, when flavour is taken into account,
the parameter $\xi$ indeed plays a role in the computation of
the baryon asymmetry.}

%%%%%%%%%%%%%%%%%%%%%%%%%%%%%%%%%%%%%%%%%%%%%%%%%%%%%%%%%%%%
\begin{figure}[t!]%%%%%%%%%%%%%%%%%%%%%%%%%%%%%%%%%%%%%%%%%%%
  %%%
  \centerline{\hspace{-1.5cm}
    \scalebox{0.55}{\includegraphics{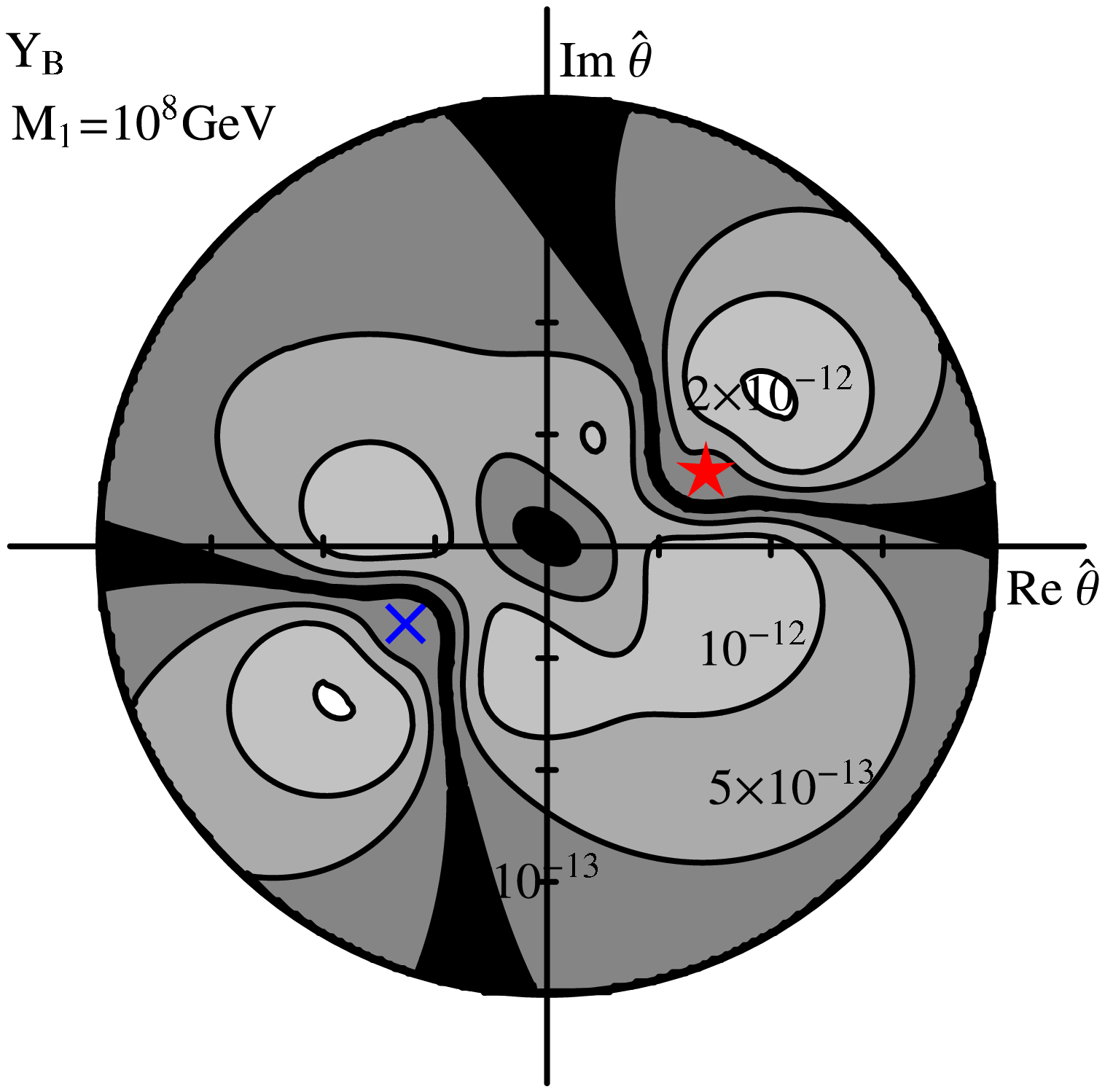}}\hspace{-1.5cm}
    \scalebox{0.55}{\includegraphics{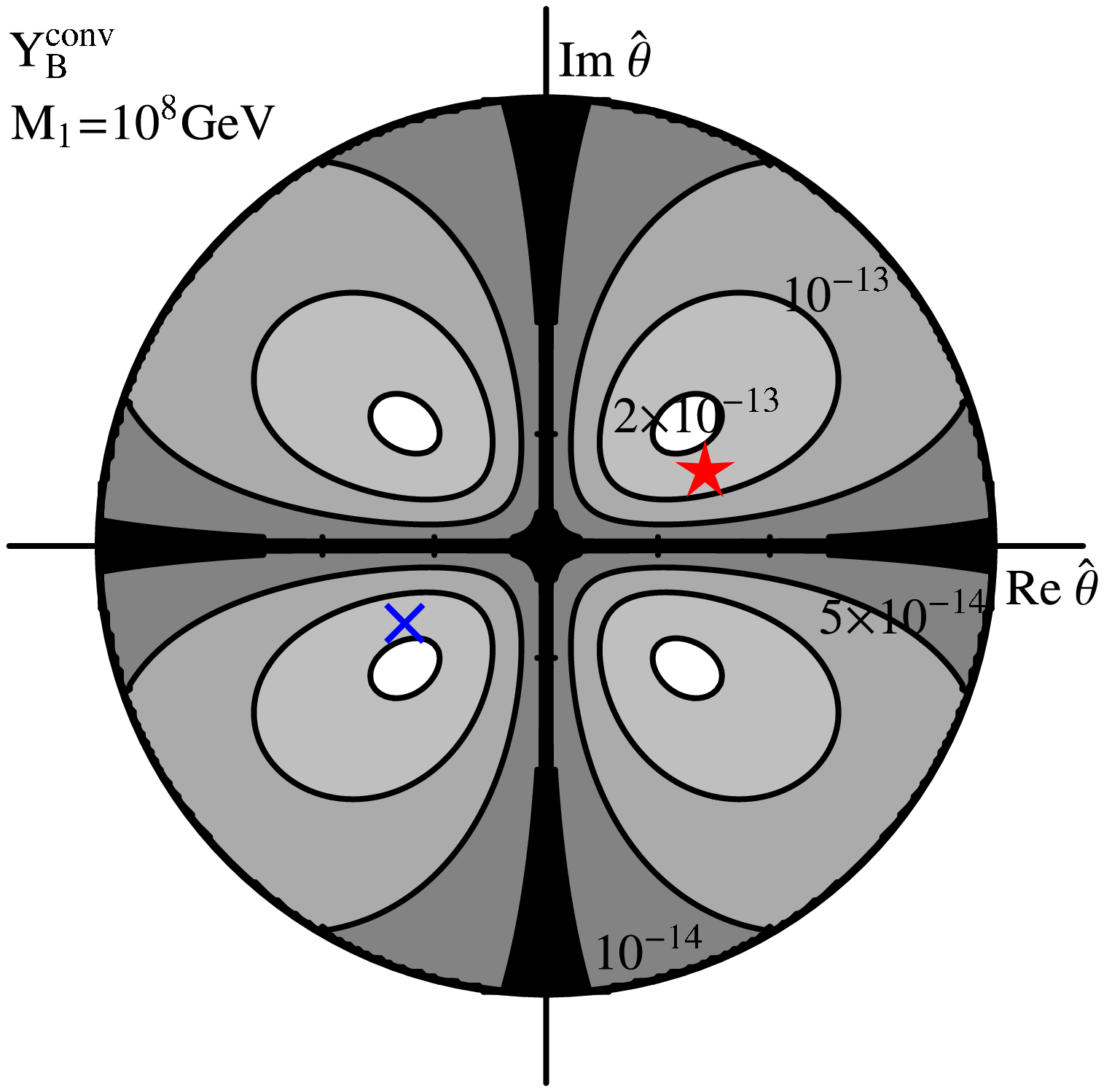}}
  }
  \centerline{\hspace{-1.5cm}
    \scalebox{0.55}{\includegraphics{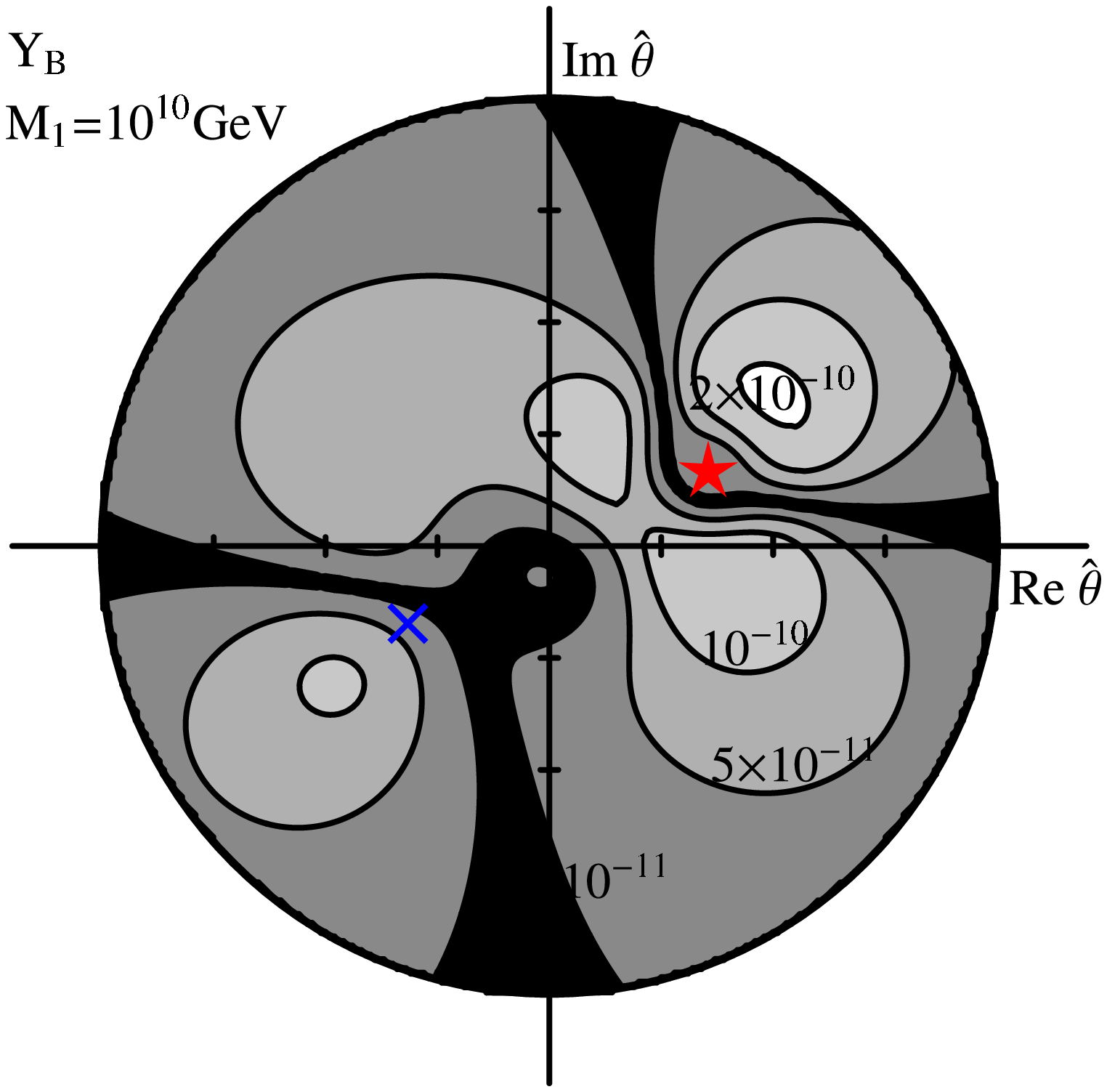}} \hspace{-1.5cm} 
    \scalebox{0.55}{\includegraphics{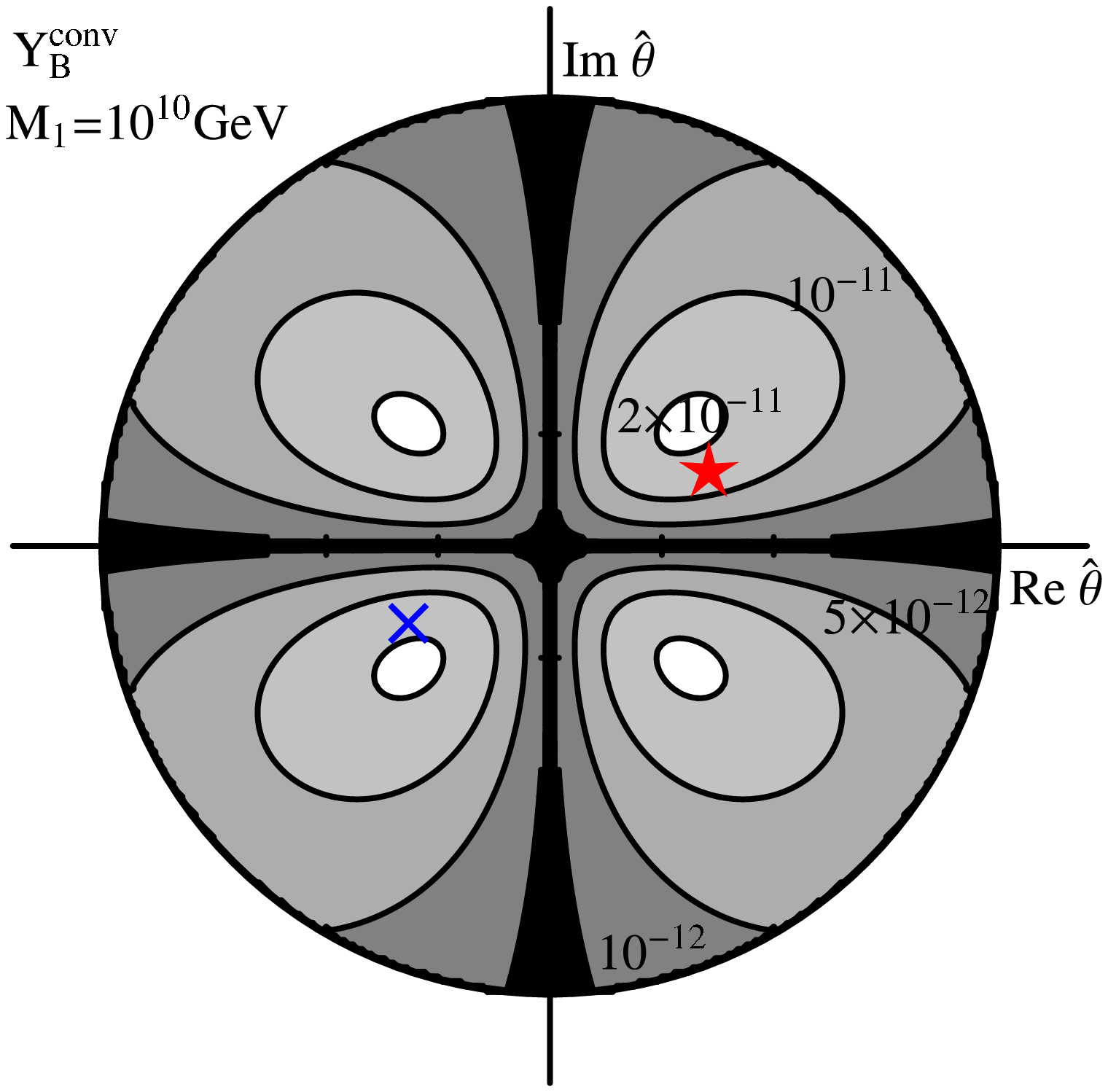}}
  } 
   
  \caption{\small Baryon asymmetry in the 2RHN model
when the neutrinos present a spectrum with normal hierarchy.
In the left plots we show the result of the calculation
that takes flavour into account, whereas in the right plots,
the result of the conventional calculation that ignores
flavour. On the top plots we take
$M_1=10^8$GeV, so that the tau and muon Yukawa couplings
are in equilibrium, while in the lower plots,
$M_1=10^{10}$GeV, so that only the tau Yukawa coupling
is in equilibrium. In these plots we have fixed $\theta_{13}=0.1$,
$\delta=\pi/4$, $\phi=\pi/3$ and the remaining neutrino
parameters to their favoured experimental values.
}
  \label{normal}
\end{figure}%%%%%%%%%%%%%%%%%%%%%%%%%%%%%%%%%%%%%%%%%%%%%%%%
%%%%%%%%%%%

The differences between the correct analysis of leptogenesis,
taking flavour into account, and the conventional analysis
are more acute along the axes ${\rm Im}\hat\theta=0$
and ${\rm Re}\hat\theta=0$, and around the values of 
$\hat\theta$ that correspond to texture zeros in the 
Yukawa coupling. The corresponding values of $\hat\theta$
are indicated in the plots with a red star, when $|\lambda_{13}|\simeq0$,
and with a  blue cross, when  $|\lambda_{12}|\simeq0$ 
(a texture zero in the (1,1)  position appears at $|\hat\theta|>1$). 
The point in the $\hat \theta$ complex plane where 
$|\lambda_{1 \alpha}|\simeq 0$ 
can be derived from eq.(\ref{yukawa-elements-norm}), 
being the result:
\bea
\tan\hat\theta^{(\alpha)}_0&\simeq &
-\xi\sqrt{\frac{m_2}{m_3}} \frac{U_{\alpha 2}^*}{U_{\alpha 3}^*}.
\eea

The reason why in the conventional analysis 
the baryon asymmetry vanishes along the axes
can be easily understood from the expression of the total 
CP asymmetry in the 2RHN model. For instance, 
for the case with normal hierarchy:
\bea
\epsilon_{1}&\simeq& \frac{3M_1}{8\pi v^2}
\frac{(m^2_3-m^2_2)\;{\rm Im} \sin^2\hat\theta}
{m_2|\cos^2\hat\theta|+m_3|\sin^2\hat\theta|}.
\label{CP-flavour-2RHN-conv}
\eea
Hence, the total CP  asymmetry vanishes
when ${\rm Im}\hat\theta=0$ ({\it i.e.} $R$ real) and when
${\rm Re}\hat\theta=0$ (since in this case
$\sin^2\hat\theta=-\sinh^2|\hat\theta|$, and $R$ is again real).
This is not necessarily the case when flavour
is properly taken into account, as can be realized from the
expressions of the flavour CP asymmetries, eq.(\ref{later}).

On the other hand, when the Yukawa coupling presents
an approximate texture zero, the difference stems
mainly from the  washing-out of the asymmetry.
When the spectrum has a normal hierarchy, the
CP asymmetry in the $\alpha$-th flavour 
is comparable to the total CP asymmetry.
However, the smallness of the interaction of the 
lightest right-handed neutrino with $\ell_\alpha$ translates into
a weak wash-out of the asymmetry, in stark contrast with the 
result of the same analysis following the conventional computation, 
where the total CP asymmetry is necessarily strongly washed-out
(recall that in the 2RHN model with normal hierarchy
$\widetilde m_1\geq \sqrt{\Delta m^2_{sol}}$, so $K>1$).
As a consequence, when $\lambda_{1\alpha}\simeq 0$, the 
actual prediction for the baryon asymmetry can be around one order
of magnitude larger than previously believed.
Interestingly enough, in realistic models 
strict texture zeros rarely appear; there are usually subleading
effects that produce small entries instead of strict 
texture zeros. To be precise, 
texture zeros normally appear in a basis where the charged lepton
Yukawa coupling and/or the right-handed neutrino mass matrix
are slightly non diagonal. Therefore, the diagonalization of these
matricial couplings to express the Lagrangian as in eq.(\ref{L})
will lift the texture zero in the neutrino Yukawa matrix, 
yielding a small entry instead. Finally, even if the texture zero 
was strict at the high energy scale, radiative corrections could 
lift them.

The differences between the computation of the baryon
asymmetry taking into account flavour or not are even
more acute in the case of a spectrum with inverted
hierarchy, as can be realized from Fig. \ref{inverted}.
As in the case of the spectrum with normal hierarchy,
the maximum differences arise along the axes
${\rm Im}\hat\theta=0$ and ${\rm Re}\hat\theta=0$, and around 
the values of $\hat\theta$ that correspond
to texture zeros in the first row of the Yukawa coupling,
also indicated in these plots with a red star and a blue cross
\footnote{Notice the proximity of both points, which is due to the
maximal atmospheric mixing. Furthermore, as $\theta_{13}$ 
approaches zero, the two points collapse into one, which 
reflects the fact that in the limit with 
$\theta_{13}=0$ and $\theta_{23}=\pi/4$,
there is an exact $\mu\leftrightarrow \tau$ symmetry,
and imposing $\lambda_{12}=0$ automatically implies $\lambda_{13}=0$.}.
The precise value
of $\hat\theta$ where $|\lambda_{1\alpha}|\simeq0$ is:
\bea
\tan\hat\theta^{(\alpha)}_0&\simeq& 
-\xi\sqrt{\frac{m_1}{m_2}} \frac{U_{\alpha 1}^*}{U_{\alpha 2}^*}.
\eea
Around this value for $\hat\theta$, the difference between the conventional
calculation (right plots) and the calculation taking into
account flavour  can be as large as three orders of magnitude
when $M_1=10^8\GeV$ (see upper left plot) or two orders of magnitude
when $M_1=10^{10}\GeV$ (see lower left plot).

%%%%%%%%%%%%%%%%%%%%%%%%%%%%%%%%%%%%%%%%%%%%%%%%%%%%%%%%%%%%
\begin{figure}[t!]%%%%%%%%%%%%%%%%%%%%%%%%%%%%%%%%%%%%%%%%%%%
  %%%zs
  \centerline{\hspace{-1.5cm}
    \scalebox{0.55}{\includegraphics{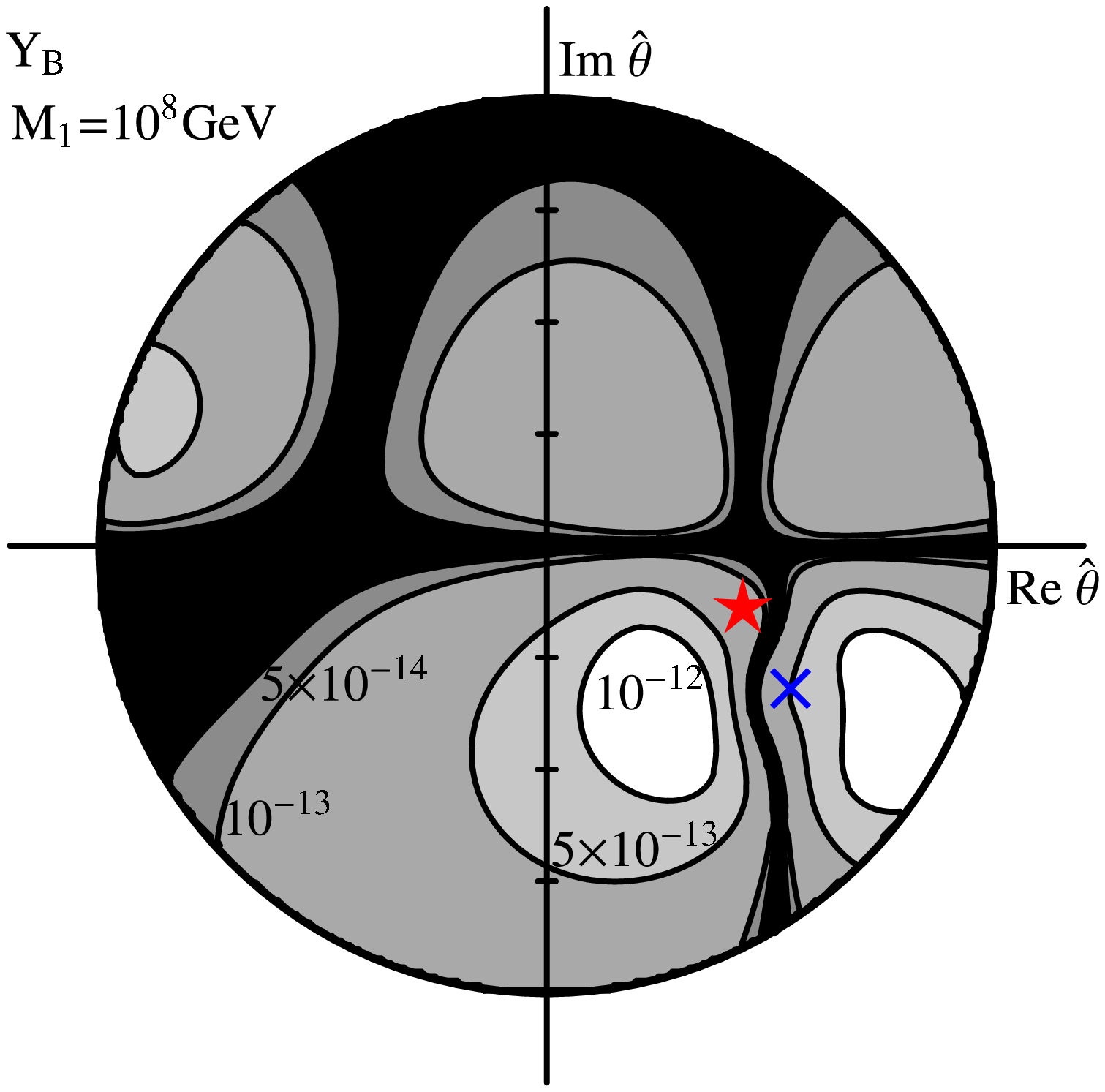}}\hspace{-1.5cm}
    ~
    \scalebox{0.55}{\includegraphics{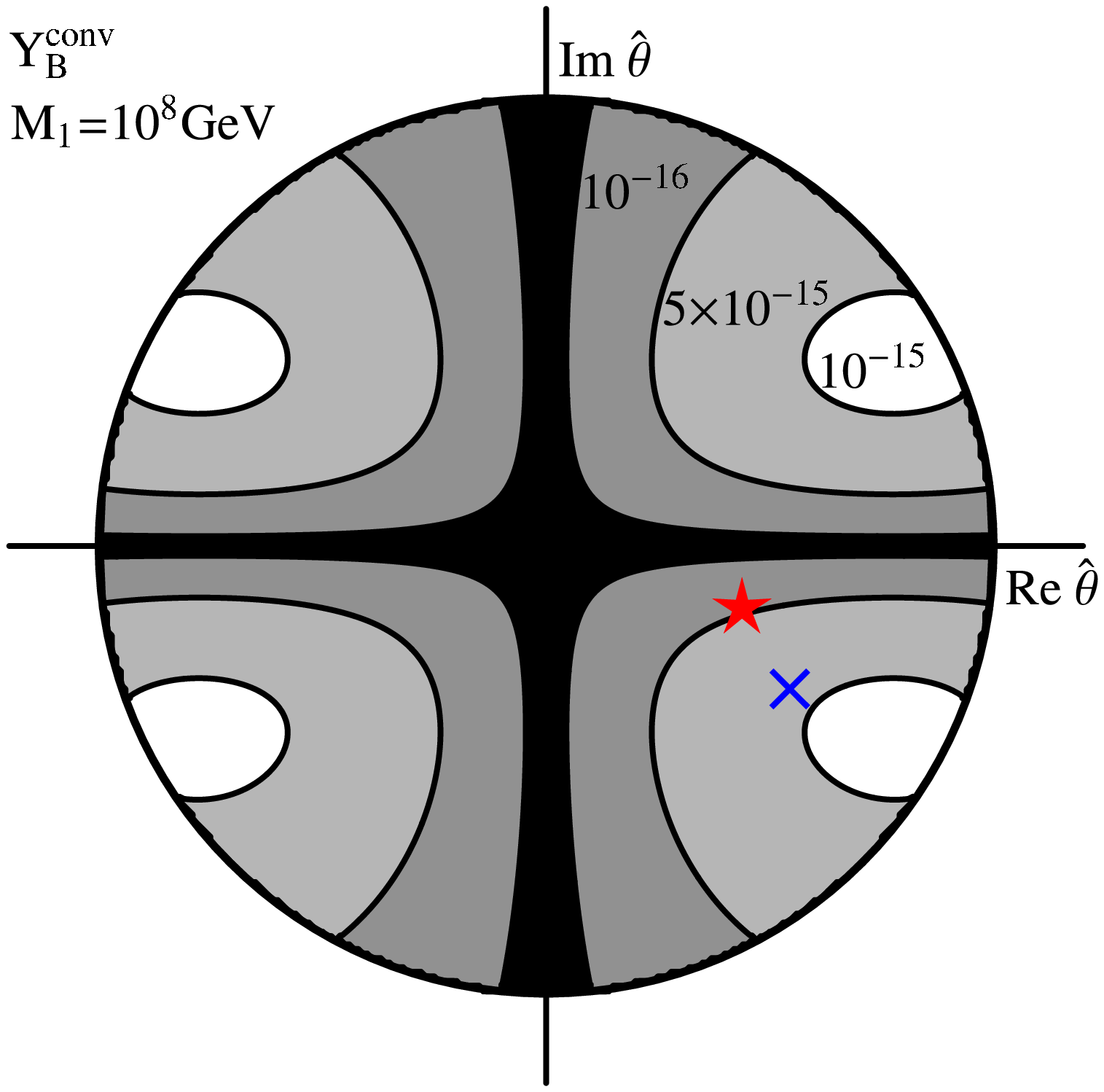}}}
  %%%
 \centerline{\hspace{-1.5cm}
   \scalebox{0.55}{\includegraphics{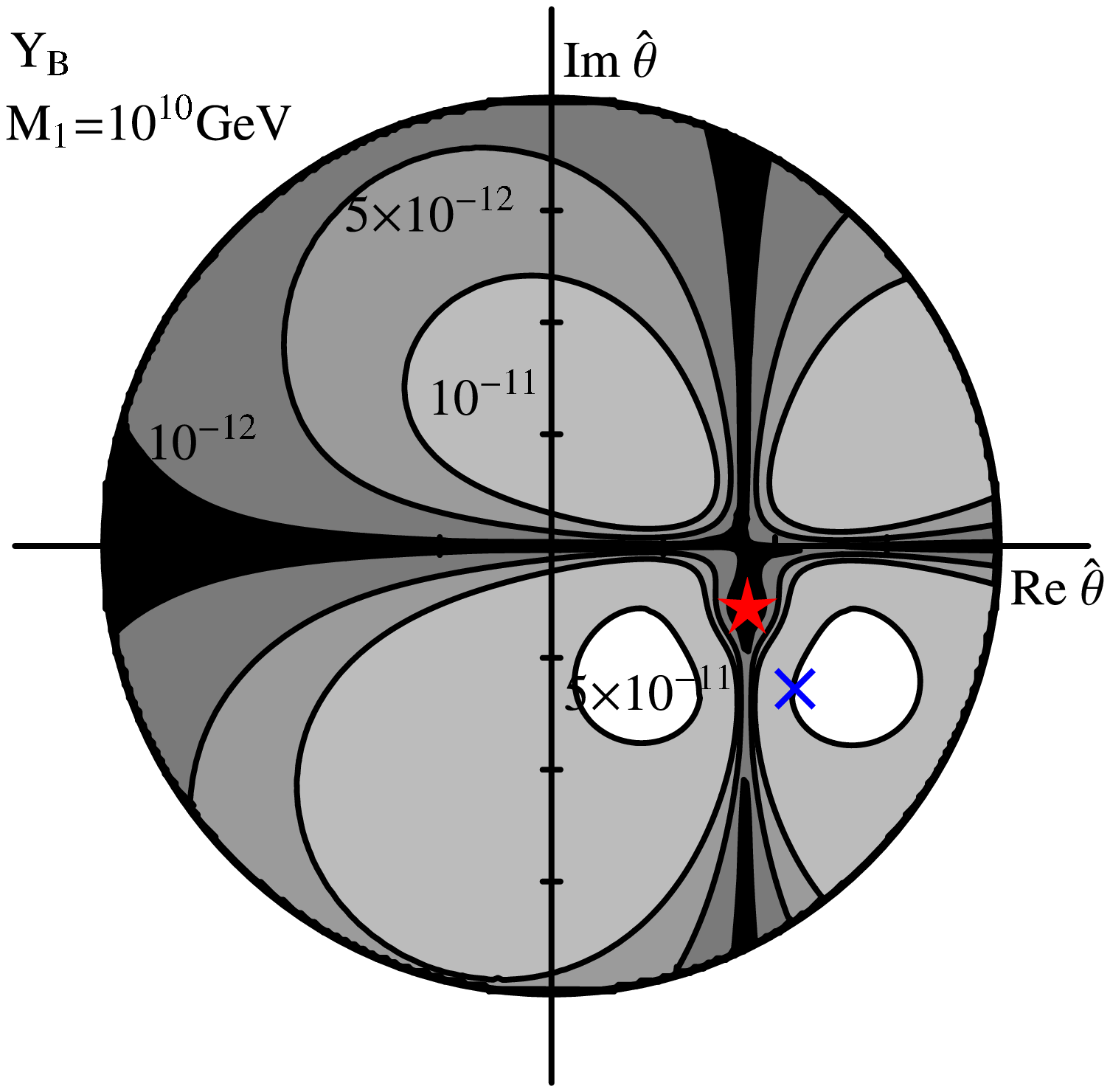}}\hspace{-1.5cm}
    ~
    \scalebox{0.55}{\includegraphics{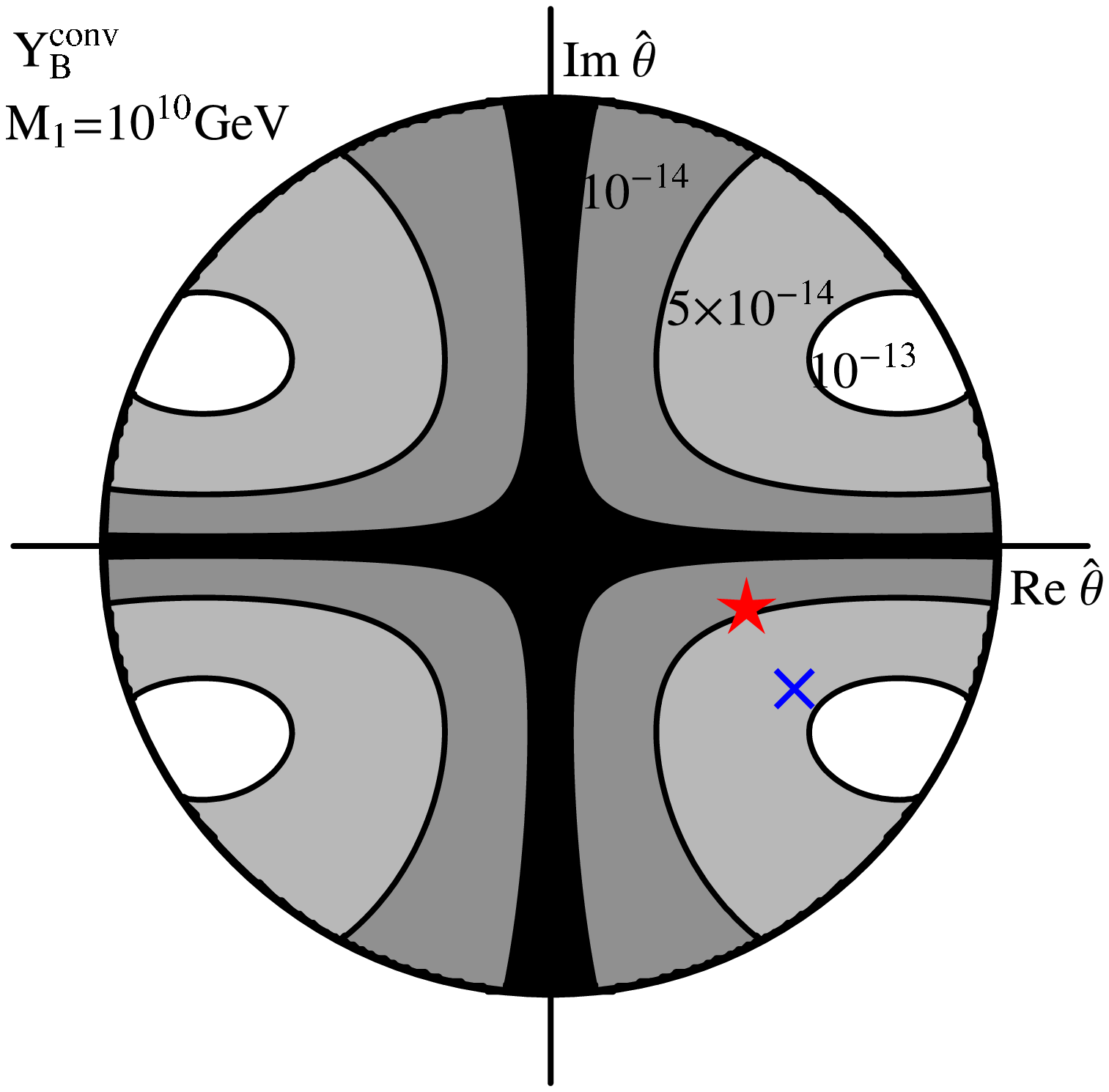}}
  }  
  \caption{\small The same as Fig.(\ref{normal}), but for
a spectrum with inverted hierarchy.}
  \label{inverted} 
\end{figure}%%%%%%%%%%%%%%%%%%%%%%%%%%%%%%%%%%%%%%%%%%%%%%%%
%%%%%%%%%%%

The reason for this huge enhancement is double. First, in the 
conventional calculation ignoring flavour, when the 
spectrum has an inverted hierarchy the total 
CP asymmetry goes as $\Delta m^2_{sol}/\sqrt{\Delta m^2_{atm}}$.
However, the flavour CP asymmetries go
as  $\sqrt{\Delta m^2_{atm}}$, therefore, the individual 
flavour CP asymmetries can be a factor
of 20 bigger than the total CP asymmetry. Secondly, the 
total lepton asymmetry computed ignoring flavour is strongly
washed-out, since $\widetilde m_1\geq \sqrt{\Delta m^2_{atm}}$.
yielding a suppressed baryon asymmetry. On the contrary,
when there is an approximate texture zero, $\lambda_{1\alpha}\simeq 0$,
the lepton asymmetry in the $\alpha$-th flavour is only weakly washed-out.
These two effects combined
are the responsible of the huge enhancement of the baryon
asymmetry when the spectrum has an inverted hierarchy and
there is an approximate texture zero in the first row
of the neutrino Yukawa coupling.

\section{Texture zeros in the two right-handed neutrino model}

In this section we would like to study carefully the predictions
for the baryon asymmetry in the case that there are approximate
texture zeros in the neutrino Yukawa coupling.
Texture zeros commonly arise in model constructions based 
on the Froggatt-Nielsen mechanism \cite{Froggatt:1978nt}. The assignment of 
different charges under an extra symmetry to particles
of different generations, translates into a Yukawa coupling
with a non-trivial structure in the effective theory, 
once the extra symmetry is spontaneously broken. 
The assignment of charges could be such that 
the resulting Yukawa coupling could have one or several 
entries which are very small compared to the others. In some
instances, the entry for the Yukawa coupling could  exactly vanish 
at the high energy scale, although these vanishing
entries are usually filled when the fields are brought to the basis
where the charged lepton Yukawa coupling and the right-handed mass
matrix are both diagonal. In addition to this, radiative effects 
can also fill these entries.

In the 2RHN model an exact texture zero in the Yukawa coupling 
fixes the value of $\tan\hat\theta$ in terms of
low energy data. Namely, if $|\lambda_{1 \beta}|=0$,
\bea
\tan\hat\theta^{(\beta)}_0&\simeq &
-\xi\sqrt{\frac{m_2}{m_3}} \frac{U_{\beta 2}^*}{U_{\beta 3}^*} ~~~~\rm{(normal~hierarchy)},\\
\tan\hat\theta^{(\beta)}_0&\simeq& 
-\xi\sqrt{\frac{m_1}{m_2}} \frac{U_{\beta 1}^*}{U_{\beta 2}^*} ~~~~\rm{(inverted~hierarchy)}.
\eea

By perturbing the Yukawa coupling around this value 
of $\hat\theta^{(\beta)}_0$ it is possible to lift
the texture zero, while still reproducing the observed
masses and mixing angles. The $R$-matrices that yield
a viable Yukawa coupling with an approximate texture zero are:
\begin{equation}
R=\left( 
\begin{array}{ccc}
0 & \cos \hat\theta^{(\beta)}_0 & \xi \sin \hat\theta^{(\beta)}_0 \\ 
0 & -\sin \hat\theta^{(\beta)}_0 & \xi \cos \hat\theta^{(\beta)}_0  
\end{array} \right)-\rho e^{i \omega}
\left( 
\begin{array}{ccc}
0 & \sin \hat\theta^{(\beta)}_0 &-\xi \cos \hat\theta^{(\beta)}_0 \\ 
0 &  \cos \hat\theta^{(\beta)}_0 & \xi \sin \hat\theta^{(\beta)}_0  
\end{array} \right)~~~
 (\rm{normal~hierarchy}),
\end{equation}
\begin{equation}
R=\left( 
\begin{array}{ccc}
\cos \hat\theta^{(\beta)}_0 & \xi \sin \hat\theta^{(\beta)}_0 & 0\\  
-\sin \hat\theta^{(\beta)}_0 & \xi \cos \hat\theta^{(\beta)}_0  & 0
\end{array} \right) -\rho e^{i \omega}
\left( 
\begin{array}{ccc}
 \sin \hat\theta^{(\beta)}_0 &-\xi \cos \hat\theta^{(\beta)}_0 &0\\ 
  \cos \hat\theta^{(\beta)}_0 & \xi \sin \hat\theta^{(\beta)}_0 &0
\end{array} \right)~~~(\rm{inverted~hierarchy}),  \label{R2x3-TZ}
\end{equation}
where $\rho e^{i \omega}$ parametrizes
the departure from the strict texture zero in the 
$\hat\theta$-parameter space, {\it i.e.} $\hat\theta=\hat\theta^{(\beta)}_0 +
\rho e^{i\omega}$. In the case with normal hierarchy
the Yukawa couplings explicitly read,
\bea
\lambda_{1\alpha}&\simeq \sqrt{\frac{M_1}{{\cal M}_{\beta\beta}}}
\left[ \sqrt{m_2 m_3} \left(U_{\alpha2}^*U^*_{\beta 3} - 
U^*_{\beta 2} U_{\alpha3}^*\right)+
\xi \rho e^{i \omega} {\cal M}_{\alpha\beta}\right]/v, \nonumber \\
\lambda_{2\alpha}&\simeq \sqrt{\frac{M_2}{{\cal M}_{\beta\beta}}}
\left[ \xi {\cal M}_{\alpha\beta} -\sqrt{m_2 m_3}
\rho e^{i \omega}\left(U_{\alpha2}^*U^*_{\beta 3} - 
U^*_{\beta 2} U_{\alpha3}^*\right)\right] /v,
\eea
while in the case with inverted hierarchy,
\bea
\lambda_{1\alpha}&\simeq \sqrt{\frac{M_1}{{\cal M}_{\beta\beta}}}
\left[ \sqrt{m_1 m_2} \left(U_{\alpha1}^*U^*_{\beta 2} - 
U^*_{\beta 1} U_{\alpha2}^*\right)+
\xi \rho e^{i \omega} {\cal M}_{\alpha\beta}\right]/v, \nonumber \\
\lambda_{2\alpha}&\simeq \sqrt{\frac{M_2}{{\cal M}_{\beta\beta}}}
\left[ \xi {\cal M}_{\alpha\beta} -\sqrt{m_1 m_2}
\rho e^{i \omega}\left(U_{\alpha1}^*U^*_{\beta 2} - 
U^*_{\beta 1} U_{\alpha2}^*\right)\right] /v.
\eea

These expressions can be substituted in eqs.(\ref{later}) 
and eq.(\ref{alphabeta}) to derive the CP asymmetries and the 
washout factors in the $\alpha$-th flavour, in the case
that $|\lambda_{1\beta}|\simeq0$.  Expanding for small values of
$\rho$ and keeping the lowest order terms we obtain for
the case with normal hierarchy the following CP flavour asymmetries:
\bea
\epsilon_{\alpha\alpha}&\simeq& -\frac{3M_1 m_3}{8\pi v^2} 
\frac{1}{|U_{\beta2}|^2+|U_{\beta3}|^2}
{\rm Im}\left[e^{i\phi/2}
\frac{{\cal M}^*_{\beta\beta}}{|{\cal M}_{\beta\beta}|}\sum_{\gamma}
\epsilon_{\alpha\beta\gamma}\left(U_{\alpha_3}U^*_{\beta2}-\frac{m_2}{m_3}
U_{\alpha_2}U^*_{\beta3}\right)U_{\gamma 1}\right]~{\rm if}~\alpha\neq\beta,
 \nonumber \\
\epsilon_{\beta\beta}&\simeq& -\frac{3M_1 m_3}{8\pi v^2} 
\frac{|{\cal M}_{\beta\beta}|}{\sqrt{m_2 m_3}}
\frac{\xi  \rho }{|U_{\beta2}|^2+|U_{\beta3}|^2}
{\rm Im}\left[U^*_{\beta 2}U_{\beta_3}
\left(e^{i\omega}+\frac{m_2}{m_3}e^{-i\omega}\right)\right],
\label{asymmetries-normal}
\eea
and the following washout factors:
\bea
K_{\alpha\alpha}&\simeq& \frac{m_2 m_3}{\tilde{m}^* |{\cal M}_{\beta\beta}|}
\left|\sum_{\gamma}\epsilon_{\alpha\beta\gamma}U_{\gamma1}\right|^2
~{\rm if}~\alpha\neq\beta, \nonumber\\
K_{\beta\beta}&\simeq& \frac{ |{\cal M}_{\beta\beta}|}{\tilde{m}^*} \rho^2.
\label{Ks-normal}
\eea

The corresponding formulas for the case with inverted hierarchy are:
\bea
\epsilon_{\alpha\alpha}&\simeq& -\frac{3M_1 m_2}{8\pi v^2} 
\frac{1}{|U_{\beta1}|^2+|U_{\beta2}|^2}
{\rm Im}\left[e^{i\phi/2}
\frac{{\cal M}^*_{\beta\beta}}{|{\cal M}_{\beta\beta}|}\sum_{\gamma}
\epsilon_{\alpha\beta\gamma}\left(U_{\alpha_2}U^*_{\beta1}-\frac{m_1}{m_2}
U_{\alpha_1}U^*_{\beta2}\right)U_{\gamma 3}\right]~{\rm if}~\alpha\neq\beta,
 \nonumber \\
\epsilon_{\beta\beta}&\simeq& -\frac{3M_1 m_2}{8\pi v^2} 
\frac{|{\cal M}_{\beta\beta}|}{\sqrt{m_1 m_2}}
\frac{\xi  \rho }{|U_{\beta1}|^2+|U_{\beta2}|^2}
{\rm Im}\left[U^*_{\beta 1}U_{\beta_2}
\left(e^{i\omega}+\frac{m_1}{m_2}e^{-i\omega}\right)\right],
\label{asymmetries-inv}
\eea
and
\bea
K_{\alpha\alpha}&\simeq& \frac{m_1 m_2}{\tilde{m}^* |{\cal M}_{\beta\beta}|}
\left|\sum_{\gamma}\epsilon_{\alpha\beta\gamma}U_{\gamma3}\right|^2,
{\rm if}~\alpha\neq\beta,  \nonumber\\
K_{\beta\beta}&\simeq& \frac{ |{\cal M}_{\beta\beta}|}{\tilde{m}^*} \rho^2.
\label{Ks-inv}
\eea

The predictions for the flavour asymmetries and the washout 
factors, and accordingly for the final baryon asymmetry,
 will depend on the low energy observables and the
``texture zero uplifting'' parameters $\rho$ and $\omega$.
In some cases, the prediction of the baryon asymmetry 
will not depend on $\rho$ and $\omega$, and therefore it
could be possible to establish a connection between leptogenesis
and low energy observables \cite{Ibarra:2003xp,connection-TZs}.
Let us analyze separately the different possibilites:

\subsection{Texture zero in the (1,1) position}

The possibility of a texture zero in the (1,1) position
is perhaps the most interesting one from the 
phenomenological point of view.
The postulates that the up and down quark matrices 
are symmetric in the first two generations and that
they present a simultaneous zero in the (1,1) position,
lead to the renown prediction for the 
Cabibbo angle $\lambda_C\simeq  \sqrt{m_d/m_s}$ \cite{Gatto:1968ss}. 
The success and robustness of this prediction may suggest
we apply the same rationale to the leptonic sector, and 
impose a texture zero in the (1,1) position of the 
neutrino Yukawa matrix, and perhaps also in the
charged lepton Yukawa coupling and the right-handed neutrino mass matrices.

The flavour CP asymmetries for the case with normal hierarchy
can be straightforwardly obtained from 
eq.(\ref{asymmetries-normal}), being the result:
\bea
\epsilon_{ee}&\simeq& \frac{3 M_1 m_3}{8\pi v^2}
\frac{|{\cal M}_{ee}|}{\sqrt{m_2 m_3}} 2 \xi  \rho 
\sin\theta_{13}\sin(\delta-\phi/2-\omega),
\nonumber\\
\epsilon_{\mu\mu}&\simeq& -\frac{3 M_1 m_3}{8\pi v^2}  \frac{\sqrt{3}}{8}
\frac{m_3}{|{\cal M}_{ee}|}\sin\theta_{13} \left[
\frac{m_2}{m_3}\sin\delta+\left(\frac{m_2}{m_3}\right)^2
\sin(\delta-\phi)+\frac{4}{\sqrt{3}}\sin\theta_{13}\sin(2\delta-\phi)\right],
\nonumber \\
\epsilon_{\tau\tau}&\simeq& \frac{3 M_1 m_3}{8\pi v^2} \frac{\sqrt{3}}{8}
\frac{m_3}{|{\cal M}_{ee}|}\sin\theta_{13} \left[
\frac{m_2}{m_3}\sin\delta+\left(\frac{m_2}{m_3}\right)^2
\sin(\delta-\phi)-\frac{4}{\sqrt{3}}\sin\theta_{13}\sin(2\delta-\phi)\right],
\nonumber \\
\label{CPasym-TZ11-norm}
\eea
where ${\cal M}_{ee}\simeq m_3 e^{2i\delta}\sin^2\theta_{13}+
 e^{i\phi} m_2/4$ is the (1,1) element of the effective
neutrino mass matrix (recall that 
this matrix element is precisely the relevant one for
analyses of neutrinoless double beta decay). 

On the other hand, the washout parameters are, from eq.(\ref{Ks-normal}):
\bea
K_{ee}&\simeq&  \frac{|{\cal M}_{ee}|}{\tilde{m}^*} \rho^2 ,
\nonumber\\
K_{\mu\mu}&\simeq& \frac{1}{8}\frac{m_2 m_3}{\tilde{m}^* |{\cal M}_{ee}|}
(1-2\sqrt{3}\sin\theta_{13}\cos\delta),
 \nonumber\\
K_{\tau\tau}&\simeq& \frac{1}{8}\frac{m_2 m_3}{\tilde{m}^* |{\cal M}_{ee}|}
(1+2\sqrt{3}\sin\theta_{13}\cos\delta),
\eea
with $\tilde{m}^*\simeq 3\times 10^{-3}\eV$.
In  view of the present experimental bound $\sin \theta_{13}\lsim 0.2$,
it follows that $ |{\cal M}_{ee}|\lsim m_2/4$ and accordingly
$K_{\mu\mu,\tau\tau}\gsim m_3/(2 \tilde{m}^*) \simeq 20$. 
Hence, when there is an approximate (1,1) texture zero and 
$\rho$ is sufficiently small, the muon and the tau
CP asymmetries are strongly washed-out, while the electron CP asymmetry
is only weakly washed-out.

For the case of an inverted hierarchy of neutrinos,
the flavour CP asymmetries read:
\bea
\epsilon_{ee}&\simeq& \frac{3 M_1 m_2}{8\pi v^2}
\frac{|{\cal M}_{ee}|}{\sqrt{m_1 m_2}} \xi  \rho 
\frac{\sqrt{3}}{2}\cos\omega\sin(\phi/2), \nonumber \\
\epsilon_{\mu\mu}&\simeq& -\frac{3 M_1 m_2}{8\pi v^2} 
\frac{3}{8}\left[\frac{\Delta m^2_{sol}}{\Delta m^2_{atm}}
\frac{\sin\phi}{\sqrt{10+6\cos\phi}}+\frac{\sin\theta_{13}}{\sqrt{3}}
\left(\frac{2\sin\delta+\sin(\delta-\phi)-3\sin(\delta+\phi)}
{\sqrt{10+6\cos\phi}}\right)\right], \nonumber\\
\epsilon_{\tau\tau}&\simeq& -\frac{3 M_1 m_2}{8\pi v^2} 
\frac{3}{8}\left[\frac{\Delta m^2_{sol}}{\Delta m^2_{atm}}
\frac{\sin\phi}{\sqrt{10+6\cos\phi}}-\frac{\sin\theta_{13}}{\sqrt{3}}
\left(\frac{2\sin\delta+\sin(\delta-\phi)-3\sin(\delta+\phi)}
{\sqrt{10+6\cos\phi}}\right)\right], \nonumber \\
\label{CPasym-TZ11-inv}
\eea
and the washout-parameters,
\bea
K_{ee}&\simeq& \frac{|{\cal M}_{ee}|}{\tilde{m}^*} \rho^2, 
\nonumber\\
K_{\mu\mu}&\simeq& \frac{1}{2}\frac{m_1 m_2}{\tilde{m}^* |{\cal M}_{ee}|}, 
 \nonumber\\
K_{\tau\tau}&\simeq& \frac{1}{2}\frac{m_1 m_2}{\tilde{m}^* |{\cal M}_{ee}|}, 
\eea
where in this case ${\cal M}_{ee}\simeq (3m_1+e^{i\phi} m_2)/4$.
Therefore, it follows that $K_{\mu\mu,\tau\tau}\gsim
\sqrt{\Delta m^2_{atm}}/(2 \tilde{m}^*)$, so the muon and tau asymmetries
are strongly washed-out, whereas the electron asymmetry is
weakly washed-out.

These formulas can be straightforwardly
applied to the recipes presented in subsection 4.1.3, to go from the
flavour asymetries to the baryon asymmetry in the different
regimes for $M_1$. We find numerically that when 
$M_1\lsim10^9\GeV$ it is not possible to reproduce the 
observed baryon asymmetry; in order to reproduce the
data it is necessary $M_1\gsim10^{11}\GeV$ so that 
only the tau Yukawa interactions are in thermal equilibrium. 
If this is the case, leptogenesis only depends on observables
that are in principle measurable at low energies, since
the electron asymmetry is always negligible compared to the
muon asymmetry (recall that when  $M_1\gsim10^{10}\GeV$ the
relevant quantities to compute the baryon asymmetry
are $Y_2=Y_{ee}+Y_{\mu\mu}$ and $K_2=K_{ee}+K_{\mu\mu}$). 
Using eqs.(\ref{third}), (\ref{BAUM1large})
and taking into account that  $K_{\mu\mu}\simeq K_{\tau\tau}$,
it follows that when $\sin\theta_{13}$ is large, in the
case with normal hierarchy 
$Y_{\cal B}\propto \sin^2\theta_{13}\sin(2\delta-\phi)$,
while in the case with inverted hierarchy the relation
is more complicated (it goes roughly as the term propotional
 to $\sin\theta_{13}$ in eq.(\ref{CPasym-TZ11-inv})). 
On the other hand, when $\sin\theta_{13}$ is
small, in the case with normal hierarchy
the baryon asymmetry is very suppressed (it goes as 
$\sin^2\theta_{13}$), while in the case with inverted
hierarchy $Y_{\cal B}\propto \sin\phi/\sqrt{10+6\cos\phi}$,
being in this case the asymmetry suppressed by 
$\Delta m^2_{sol}/\Delta m^2_{atm}$.

\subsection{Texture zero in the (1,2) position}

In the case that there is an approximate texture zero 
in the (1,2) position of the neutrino Yukawa matrix, 
the flavour CP asymmetries read:
\bea
\epsilon_{ee}&\simeq& \frac{3 M_1 m_3}{8\pi v^2} 
\frac{\sqrt{3}}{7}\left[\sin\theta_{13}\sin(\delta-\phi)
-\frac{\sqrt{3}}{4}\frac{m^2_2}{m^2_3}\sin\phi\right],
\nonumber \\
\epsilon_{\mu\mu}&\simeq& -\frac{3 M_1 m_3}{8\pi v^2}
\sqrt{\frac{m_3}{m_2}}\frac{\sqrt{3}}{7} \xi  \rho 
\sin(\phi/2+\omega)\nonumber, \\
\epsilon_{\tau\tau}&\simeq& \frac{3 M_1 m_3}{8\pi v^2} 
\frac{3}{7}\left[\sin\phi+\frac{\sin\theta_{13}}{\sqrt{3}}\sin(\delta-\phi)
\right].
\label{CPasym-TZ12-norm}
\eea
On the other hand, the washout parameters read:
\bea
K_{ee}&\simeq& \frac{1}{4}\frac{m_2}{\tilde{m}^*}
(1-2\sqrt{3}\sin\theta_{13}\cos\delta) ,
\nonumber\\
K_{\mu\mu}&\simeq&\frac{1}{2}\frac{m_3}{\tilde{m}^*}\rho^2 ,
\nonumber \\
K_{\tau\tau}&\simeq& \frac{3}{2}\frac{m_2}{\tilde{m}^*} .
\eea
Therefore, the electron asymmetry and the muon asymmetries
are only weakly washed-out, while the tau asymmetry is 
strongly washed-out (when $M_1\gsim 10^{10}$GeV, 
$Y_2$ would be weakly washed-out
and $Y_{\tau\tau}$, strongly washed-out).

We find that when there is an approximate
texture zero in the (1,2) position,  $M_1\gsim  10^{10}\GeV$ 
is necessary in order to reproduce the observed 
baryon asymmetry. In the strict texture zero limit, the baryon
asymmetry is dominated by the tau lepton asymmetry and
therefore there is a well defined connection between leptogenesis and
low energy observables, $Y_{\cal B}\propto\sin\phi$. Despite
this connection it becomes more diffuse as we depart from the texture
zero limit, the connection still holds in the region in
the vicinity of the (1,2) texture zero where the 
baryon asymmetry is enhanced (see Fig.(\ref{normal}) lower left plot).

We find a similar behaviour when the spectrum presents
an inverted hierarchy. In this case the flavour
asymmetries are:
\bea
\epsilon_{ee}&\simeq& -\frac{3 M_1 m_2}{8\pi v^2}\frac{3}{4}
\left[\frac{\Delta m^2_{sol}}{\Delta m^2_{atm}}
\frac{\sin\phi}{\sqrt{10+6\cos\phi}}+\frac{\sin\theta_{13}}{\sqrt{3}}
\left(\frac{2\sin\delta+\sin(\delta-\phi)-3\sin(\delta+\phi)}
{\sqrt{10+6\cos\phi}}\right)\right], \nonumber \\
\epsilon_{\mu\mu}&\simeq& -\frac{3 M_1 m_2}{8\pi v^2}
\sqrt{\frac{3}{2}}\frac{\xi  \rho}{8} \sqrt{5+3\cos\phi} 
\sin(\phi/2)\cos\omega, \nonumber \\
\epsilon_{\tau\tau}&\simeq& \frac{3 M_1 m_2}{8\pi v^2}
\frac{\sqrt{3}}{4}\sin\theta_{13}
\left[\frac{2\sin\delta+\sin(\delta-\phi)-3\sin(\delta+\phi)}
{\sqrt{10+6\cos\phi}}\right],
\label{CPasym-TZ12-inv}
\eea
and the washout factors,
\bea
K_{ee}&\simeq&\frac{1}{2} \frac{m_1 m_2}{\tilde{m}^* |{\cal M}_{\mu\mu}|} , 
\nonumber\\
K_{\mu\mu}&\simeq& \frac{|{\cal M}_{\mu\mu}|}{\tilde{m}^*} \rho^2, \nonumber\\
K_{\tau\tau}&\simeq&  \frac{m_1 m_2}{\tilde{m}^* |{\cal M}_{\mu\mu}|}\sin^2\theta_{13},
\eea
with  ${\cal M}_{\mu\mu}\simeq m_1/8+3/8m_2 e^{i \phi}$.
Using that $|{\cal M}_{\mu\mu}|\lsim m_2/2$, we find
that the electron asymmetry is necessarily strongly washed-out,
whereas the muon and the tau asymmetries are only weakly washed-out
(when $M_1\gsim 10^{10}$GeV, $Y_2$ would be strongly washed-out
and $Y_{\tau\tau}$, weakly washed-out).

For the case with inverted hierarchy,
we require again $M_1\gsim  10^{10}$GeV to reproduce the
observed asymmetry (or even larger, when $\sin\theta_{13}$
is small). The baryon asymmetry is in this case also dominated
by the tau asymmetry, except when  $\sin\theta_{13}$ is very small.
In the case that the baryon asymmetry is dominated by the tau
asymmetry, although there exists a connection between leptogenesis
and low energy observables, this connection is too complicated
to be of any practical use. On the other hand, when
 $\sin\theta_{13}$ is very small, there is no relation whatsoever,
since leptogenesis would depend on the unobservable parameters
$\rho$ and $\omega$.

\subsection{Texture zero in the (1,3) position}

Finally, in the case that the texture zero appears in the (1,3)
position, the flavour CP asymmetries are:
\bea
\epsilon_{ee}&\simeq& -\frac{3 M_1 m_3}{8\pi v^2} 
\frac{\sqrt{3}}{7}\left[\sin\theta_{13}\sin(\delta-\phi)
+\frac{\sqrt{3}}{4}\frac{m^2_2}{m^2_3}\sin\phi\right],\nonumber\\
\epsilon_{\mu\mu}&\simeq& \frac{3 M_1 m_3}{8\pi v^2} 
\frac{3}{7}\left[\sin\phi-\frac{\sin\theta_{13}}{\sqrt{3}}\sin(\delta-\phi)
\right],\nonumber\\
\epsilon_{\tau\tau}&\simeq& \frac{3 M_1 m_3}{8\pi v^2} 
\sqrt{\frac{m_3}{m_2}} \frac{\sqrt{3}}{7} \xi  \rho 
\sin(\phi/2+\omega).
\label{CPasym-TZ13-norm}
\eea
On the other hand, the washout parameters read:
\bea
K_{ee}&\simeq& \frac{1}{4}\frac{m_2}{\tilde{m}^*}
(1+2\sqrt{3}\sin\theta_{13}\cos\delta) , 
\nonumber\\
K_{\mu\mu}&\simeq& \frac{3}{2}\frac{m_2}{\tilde{m}^*} ,
\nonumber\\
K_{\tau\tau}&\simeq&\frac{1}{2}\frac{m_3}{\tilde{m}^*}\rho^2 .
\eea
Therefore, in this case, the electron and the tau asymmetries
are weakly washed-out, and the muon asymmetry, strongly
washed-out. On the other hand, for $M_1\gsim 10^{9}$GeV the 
relevant quantity to estimate the washout is $K_2=K_{ee}+
K_{\mu\mu}>1$, so in this regime $Y_2$ is strongly washed out 
 and $Y_{\tau\tau}$ is weakly washed out. 

Similarly to the case of the (1,2) texture zero, this case
requires $M_1\gsim  10^{10}\GeV$ to reproduce the observations.
Furthermore, the baryon asymmetry in the vicinity of the
texture zero is dominated by the muon asymmetry and hence depends
mainly on $\sin\phi$. This behaviour occurs in particular,
in the region where the baryon asymmetry is enhanced
in Fig.(\ref{normal}), lower left plot.

The case with inverted hierarchy presents some qualitative
differences with respect to the case with normal hierarchy.
When neutrinos have an inverted hierarchy, the flavour
CP asymmetries read:
\bea
\epsilon_{ee}&\simeq& -\frac{3 M_1 m_2}{8\pi v^2} 
\frac{3}{4}\left[\frac{\Delta m^2_{sol}}{\Delta m^2_{atm}}
\frac{\sin\phi}{\sqrt{10+6\cos\phi}}-\frac{\sin\theta_{13}}{\sqrt{3}}
\left(\frac{2\sin\delta+\sin(\delta-\phi)-3\sin(\delta+\phi)}
{\sqrt{10+6\cos\phi}}\right)\right], \nonumber \\
\epsilon_{\mu\mu}&\simeq&- \frac{3 M_1 m_2}{8\pi v^2}
\frac{\sqrt{3}}{4}\sin\theta_{13}
\left[\frac{2\sin\delta+\sin(\delta-\phi)-3\sin(\delta+\phi)}
{\sqrt{10+6\cos\phi}}\right], \nonumber \\
\epsilon_{\tau\tau}&\simeq& -\frac{3 M_1 m_2}{8\pi v^2}
\sqrt{\frac{3}{2}}\frac{\xi  \rho}{8} \sqrt{5+3\cos\phi} 
\sin(\phi/2)\cos\omega,
\label{CPasym-TZ13-inv}
\eea
and the washout parameters,
\bea
K_{ee}&\simeq& \frac{1}{2} \frac{m_1 m_2}{\tilde{m}^* |{\cal M}_{\tau\tau}|}, 
\nonumber\\
K_{\mu\mu}&\simeq&  \frac{m_1 m_2}{\tilde{m}^* |{\cal M}_{\tau\tau}|}
\sin^2\theta_{13},
\nonumber\\
K_{\tau\tau}&\simeq& \frac{|{\cal M}_{\tau\tau}|}{\tilde{m}^*} \rho^2 ,
\eea
with   ${\cal M}_{\tau\tau}\simeq m_1/8+3/8m_2 e^{i \phi}$.
As for the case with the (1,2) texture zero, we find that
the electron asymmetry is necessarily strongly washed-out,
whereas the muon and the tau asymmetries are only weakly washed-out
(also, in the regime where only the tau Yukawa interactions are
in equilibrium, $Y_2$ would be strongly washed-out
and $Y_{\tau\tau}$, weakly washed-out).

It is important to note that all the flavour asymmetries
in eq.(\ref{CPasym-TZ13-inv})
have a suppression factor, with different origins. As a consequence,
in the limit of the  strict texture zero, the resulting
baryon asymmetry is very small and  $M_1\gsim 5\times 10^{12}$GeV 
would be necessary to accommodate observationl data. However,
as we depart from the texture zero limit, we find a huge
enhancement of the baryon asymmetry, that can allow right-handed neutrino
masses as low as $M_1\sim 10^{10}$GeV, independently of the value of $\sin\theta_{13}$.
The reason is that the tau CP asymmetry can become less suppressed,
and at the same time the resulting tau lepton asymmetry can be sizable
since the tau asymmetry is only weakly washed-out 
(on the contrary, $\epsilon_2=\epsilon_{ee}+\epsilon_{\mu\mu}$
could be comparable
to $\epsilon_{\tau\tau}$, but is strongly washed-out). As a result,
the baryon asymmetry is dominated by the tau asymmetry and hence
any connection between leptogenesis and low energy observables
is lost in this region of enhanced baryon asymmetry.

\section{The case of $R$ real}

In Section 5 it was discussed that there are two situations
where the differences between the computation of the baryon
asymmetry taking into account flavour or not are maximal, 
namely when there is an approximate texture zero in the
neutrino Yukawa matrix, and when the matrix $R$ is real. 
The case with $R$ real physically corresponds to the class
of models where CP is an exact symmetry in the
right-handed neutrino sector. The reason for this 
can be more easily understood working in the basis where the 
charged lepton Yukawa coupling and the 
right-handed mass matrix are diagonal, so that 
the neutrino Yukawa matrix is the only
coupling in the leptonic Lagrangian that violates CP. 
More specifically, the neutrino Yukawa coupling
can be written in its singular value decomposition, 
$\lambda= V^{\dagger}_R {\rm Diag}(\lambda_1,\lambda_2,\lambda_3) V_L$.
Hence, the CP violation in the right-handed neutrino sector 
is encoded in the  phases in $V_R$, that can be extracted from
diagonalizing the combination $\lambda \lambda^{\dagger}=  V^{\dagger}_R
{\rm Diag}(\lambda^2_1,\lambda^2_2,\lambda^2_3) V_R$. On the other
hand, using the parametrization of the Yukawa coupling in
eq.(\ref{yukawa}), this same combination of matrices 
can be written as 
 $\lambda \lambda^{\dagger}=M^{1/2} R m R^{\dagger}M^{1/2}/v^2$.
Comparing the two expressions it is apparent that $R$ is real
if and only if $V_R$ is real, {\it i.e.} when there is no CP violation
in the right-handed sector.\footnote{Furthermore, it can
be checked that there is mixing in $V_R$ if and only if there
is mixing in $R$, and that mixing in any $2\times 2$ block in 
$R$ translates into mixing in the same block of the matrix $V_R$.}

In this limit the flavour CP asymmetries and the baryon asymmetry
depend exclusively on the phases of the left-handed sector, that are
in turn uniquely determined by the low energy phases. Consequently, 
in this limit the leptogenesis mechanism is tightly connected
to the low energy phases. This connection is more apparent
from the expression of the flavour CP asymmetries in the
parametrization eq.(\ref{yukawa}):
\bea
\epsilon_{\alpha\alpha}&\simeq& \frac{3M_1}{8\pi v^2}
\frac{{\rm Im}\left(\sum_\beta\sqrt{m_\beta}R_{1\beta} U^*_{\alpha\beta}\right)
\left( \sum_\beta \sqrt{m^3_\beta}R_{1\beta} U_{\alpha\beta}\right)}
{\sum_\beta m_\beta R^2_{1\beta}} \nonumber \\
&=& \frac{3M_1}{8\pi v^2} 
\frac{\sum_\beta \sum_{\gamma>\beta}\sqrt{m_\beta m_\gamma}
(m_\gamma-m_\beta)R_{1\beta}R_{1\gamma}
{\rm Im}U^*_{\alpha\beta}U_{\alpha\gamma}}{\sum_\beta m_\beta R^2_{1\beta}}.
\eea

This discussion suggests that the observation of low energy CP
violation would constitute an important hint to the leptogenesis mechanism. 
In a general case with $R$ complex, the low energy
phases in the leptonic mixing matrix could stem from the phases
in the left-handed sector, in the right-handed sector, or 
in both sectors. In {\it any} of the cases, and barring 
unnatural cancellations, a baryon asymmetry is necessarily 
generated through the mechanism of leptogenesis,
as long as at least one of the lepton Yukawa interactions is in
equilibrium (corresponding roughly to  $M_1\lsim 10^{12}\GeV$).
This result only follows when flavour is correctly taken
into account in the Boltzmann equations. In previous analyses
of leptogenesis ignoring flavour, the observation of low energy CP violation
did not automatically imply the existence of a baryon asymmetry,
since the possibility existed that the low energy phases 
could stem exclusively from the left-handed sector 
and hence be irrelevant for leptogenesis.

\section{Conclusions}

Thermal leptogenesis is an attractive and minimal mechanism to
make the baryon asymmetry of the Universe. The asymmetry
is commonly calculated by solving a Boltzmann equation for the
total lepton asymmetry (one-flavour approximation).
In a previous  paper 
\cite{davidsonetal}
we studied the impact of lepton flavours (charged lepton Yukawa couplings)
on the Boltzmann equations (one for each lepton flavour)
and discussed the phenomenological implications for leptogenesis. 

It may be counter-intuitive that flavour matters in
leptogenesis, since Yukawa couplings are a small perturbative correction.
We have shown that flavour effects are relevant when the interaction rates
mediated by the  the charged Yukawa couplings are faster than the
typical timescale for leptogenesis.  The charged Yukawa rates
 may be dropped from the Boltzmann 
equations {\it provided} the latter are written in the flavour basis, where
the charged Yukawa couplings cannot change the flavour of the  asymmetries. 
This implies  that one should  solve  Boltzmann equations
for each flavour. 
%In the region where our approximations 
%are valid the equations are not coupled.

The final value of the baryon asymmetry depends
on the CP asymmetry in each   flavour $\alpha$ and 
on the washing out  by the lepton number $\alpha$ 
violating processes.  Taking into account these flavour 
dependent washing out factors generically enhances the baryon asymmetry
with respect to the usual one-flavour approximation,
in the limit of strong washout.

In this paper we have provided analytical approximations for the
final baryon asymmetry with flavours accounted for.  These
depend on the temperature of leptogenesis, and can be
obtained following the procedure of section \ref{recipe1},
or of the end of section \ref{tau}.  We also  
included  CP violation in the $\Delta L=1$ scatterings relevant 
for $N_1$ production. 

In the two right-handed neutrino (2RHN) model,
we have compared our  results obtained with flavoured 
Boltzmann equations against the
usual one-flavour approximation, 
to illustrate the big impact that flavour has on leptogenesis. 
We have found that there are two situations where the differences
between the treatment of leptogenesis taking flavour properly
into account and previous analyses, that ignored flavour, are maximal.
The first one arises when the neutrino Yukawa coupling
present approximate texture zeros in the first row, so that 
the  CP asymmetry in that flavour is only weakly washed-out. 
As a consequence, we have found that thermal leptogenesis in the 2RHN
model can produce the observed baryon asymmetry for masses of the lightest
right-handed neutrino smaller than previously believed, namely $10^{10}$
GeV for the case with normal hierarchy and $5\times 10^{10}$ 
GeV for the case with inverted hierarchy 
(to be compared with $10^{11}$ GeV and $10^{13}$ GeV,
respectively, from the conventional computation ignoring flavour). 
%As a consequence, we have found that the absolute
%bound on the lightest right-handed neutrino mass in order
%to accommodate the observed baryon asymmetry can be relaxed even
%by one order of magnitude, in the case that light neutrinos present
%a spectrum with normal hierarchy, or two orders of magnitude,
%in the case that they present an inverted hierarchy. 
The second
situation corresponds to the limit in which CP is an exact
symmetry in the right-handed neutrino sector. In this case, the
conventional computation would yield an exactly vanishing
baryon asymmetry, whereas the computation that takes flavour
into account could predict a sizable baryon asymmetry.

\vskip 2cm
{\bf Acknowledgements}
\vskip 0.5cm
It is a pleasure to thank E. Nardi for sharing with us his results
and for many useful discussions. We also thank L. Covi, 
G.F. Giudice, A. Pilaftsis and A. Strumia for discussions.
SD  thanks in particular A. Strumia for many
thought-provoking and
enlightening emails, and A. Notari for related results. 
AR thanks M. Passera for useful discussions. AA and 
FXJM acknowledge the 
support of the Agence 
Nationale de la Recherche ANR through the project JC05-43009, NEUPAC.

\end{document}